\documentclass[english,aps,prx,reprint,superscriptaddress,twocolumn]{revtex4-2}
\usepackage[T1]{fontenc}
\usepackage[utf8]{inputenc}
\usepackage{float}
\usepackage{tikz}
\usepackage{amsmath}
\usepackage{graphicx}
\usepackage{amssymb}
\usepackage[colorlinks=true, allcolors=blue]{hyperref}
\usepackage{etoolbox}

\AtBeginEnvironment{thebibliography}{
    \DeclareRobustCommand{\href}[2]{#2}
    \DeclareRobustCommand{\url}[1]{#1}
    \renewcommand{\doibase}[1]{}
}

\makeatletter

\providecolor{lyxadded}{rgb}{0,0,1}
\providecolor{lyxdeleted}{rgb}{1,0,0}
\usetikzlibrary{calc}
\newcommand{\lyxobjectsout}[1]{%
  \bgroup%
  \color{lyxdeleted}%
  \tikz{
    \node[inner sep=0pt,outer sep=0pt](lyxdelobj){#1};
    \draw($(lyxdelobj.south west)+(2em,.5em)$)--($(lyxdelobj.north east)-(2em,.5em)$);
  }
  \egroup%
}
\DeclareRobustCommand{\lyxdisplayobjdeleted}[4][]{%
  \ifx#4\empty\else%
     \leavevmode\\%
     \lyxobjectsout{\parbox{\linewidth}{#4}}%
  \fi%
}
\DeclareRobustCommand{\lyxudisplayobjdeleted}[4][]{%
  \ifx#4\empty\else%
     \leavevmode\\%
     \raisebox{-\belowdisplayshortskip}{%
                \lyxobjectsout{\parbox[b]{\linewidth}{#4}}}%
     \leavevmode\\%
  \fi%
}

\usepackage{booktabs}
\usepackage{amsmath}

\makeatother

\usepackage{babel}
\begin{document}
\title{%Effective anisotropic 
  Emergence of Kugel-Khomskii %Hamiltonian for 
  physics in   %quarter-hole-filled
  quarter-filled
  %bilayer Hubbard model
  bilayer correlated systems}
\author{Guijing Duan}
\affiliation{School of Physics and Beijing Key Laboratory of Opto-electronic
  Functional Materials \& Micro-nano Devices, Renmin University of China,
  Beijing 100872, China}
\author{Yunlong Wang}
\affiliation{School of Physics and Beijing Key Laboratory of Opto-electronic
  Functional Materials \& Micro-nano Devices, Renmin University of China,
  Beijing 100872, China}
\author{Zhiguang Liao}
\affiliation{School of Physics and Beijing Key Laboratory of Opto-electronic
  Functional Materials \& Micro-nano Devices, Renmin University of China,
  Beijing 100872, China}
\author{Changle Liu}
\email{liuchangle89@gmail.com}

\affiliation{School of Physics and Mechatronic Engineering, Guizhou Minzu University,
  Guiyang 550025, China }
\author{Rong Yu}
\email{rong.yu@ruc.edu.cn}

\affiliation{School of Physics and Beijing Key Laboratory of Opto-electronic
  Functional Materials \& Micro-nano Devices, Renmin University of China,
  Beijing 100872, China}
\affiliation{Key Laboratory of Quantum State Construction and Manipulation (Ministry
  of Education), Renmin University of China, Beijing, 100872, China}
\begin{abstract}
  We present a theoretical study of the low-energy physics of a quarter-hole-filled
  two-orbital
  bilayer Hubbard model motivated by transition-metal bilayer systems
  with strong orbital-selective interlayer hybridization. By explicitly
  treating the strong
  interlayer bonding of $d_{z^{2}}$ orbitals within
  a molecular orbital basis
  and projecting out high-energy
  electronic states, we
  derive a low-energy effective
  Kugel-Khomskii Hamiltonian describing
  the interplay between electron spin and emergent layer pseudospin
  degrees of freedom.
  We map out a rich ground state phase diagram featuring diverse
  spin and charge ordered states. These include
  conventional
  ferromagnetic and antiferromagnetic phases with layer staggered
  charge densities,
  a layer-coherent phase characterized by spontaneous interlayer
  quantum coherence, and a novel maximally spin-layer-entangled phase
  with a hidden composite spin-layer order.
  We show that this exotic hidden ordered phase arises from the spontaneous
  breaking of an emergent $O(4)$
  symmetry down to a $O(3)$, manifesting a unique excitation
  spectrum with three entangled gapless Goldstone modes. Our results
  uncover a geometrically driven mechanism for realizing composite entanglement
  in strongly correlated bilayer systems and provide a concrete theoretical
  framework relevant to bilayer nickelate superconductors and other
  multi-component
  correlated materials.
\end{abstract}
\maketitle

\section{introduction}

Strongly correlated electron systems with multiple internal degrees
of freedom provide a fertile ground for
realizing exotic quantum phases
beyond the paradigms of conventional electronic orders
\cite{paschen2021quantum,schaffer2016recent,dagotto2005}. In addition to the
charge and spin degrees of freedom, orbital, layer, and valley indices
often play
an essential role in shaping the low-energy physics of a wide range
of materials, including transition-metal
compounds \cite{TMD2011,zhang2020flat,chenKK}, moir\'e superlattices
\cite{liu2021orbital,dai2016twisted,kariyado2023twisted,zhang2018lowest},
and ultracold atomic systems \cite{chenwu2024,gorshkov2010two}. In such multi-component settings, these
internal degrees of freedom are not merely passive labels but actively
participate in collective phenomena, giving rise to intertwined electronic
orders
and novel forms of quantum entanglement.

A paradigmatic framework for exploring such physics is provided by
Kugel-Khomskii type models \cite{KKM,streltsov2017orbital}, where spin and
orbital degrees
of freedom are coupled through exchange interactions generated by
virtual charge fluctuations. Traditionally, most studies of these
models focus on phases characterized by decoupled spin and orbital
orders, such as
antiferromagnetic or ferromagnetic order
accompanied by ferro-
or antiferro-orbital order \cite{brzezicki2011entangled, khaliullin1997spin}.
Even in systems with enlarged symmetries,
such as $SU(4)$-symmetric models \cite{kugel2015spin}, the emphasis has largely been
placed on symmetry enhancement, and exotic quantum liquid behavior \cite{wang2009z,calvera2021theory,zhang2024variational,jin2023twisting}.
By contrast, the structure and consequences of local spin-orbital
entanglement itself have received comparatively limited attention \cite{brzezicki2013exotic,gotfryd2020spin}.

Bilayer and multilayer quantum
materials have recently emerged as
a central theme in condensed matter physics, driven by advances ranging
from twisted moir\'e superlattices \cite{carr2020electronic,cao2020tunable,cao2016superlattice,chu2020review} to the newly discovered high-$T_{c}$ superconductors
in bilayer nickelates \cite{Sun_Nature_2023, arXiv_2407_05681Cheng,li2025identification}.
These platforms are particularly intriguing
because the layer index introduces a synthetic and tunable degree
of freedom that can compete or cooperate with spin and charge dynamics.
While much of the current
research has been focused on band engineering
and Fermiology \cite{ShilenkoLeonov_PRB_2023,LechermannEremin_PRB_2023,
  ZhangDagotto_PRB_2023,CaoYang_PRB_2024,OuyangLu_PRB_2024,
  RyeeWehling_PRL_2024,TianLu_PRB_2024,liao2412}, the interplay between the
interlayer geometry and
the orbital character of electrons offers a distinct and less explored
route toward unconventional correlated states.

A particularly compelling realization of correlated bilayer
system
is provided by the recently discovered bilayer nickelate superconductors,
such as La$_{3}$Ni$_{2}$O$_{7}$. These materials consist of NiO
bilayers separated by insulating spacer layers and exhibit
high-temperature superconductivity
under pressure. Owing to the multi-orbital nature of the Ni 3$d$
manifold and the pronounced structural anisotropy, bilayer nickelates
naturally host strong electronic correlations together with substantial
interlayer coupling effects. First-principles and spectroscopic measurements
have highlighted the crucial role of orbital-dependent hybridization,
with the $d_{z^{2}}$ orbital experiencing significant interlayer
bonding\textendash antibonding splitting, while the hopping between the
$d_{x^{2}-y^{2}}$
orbitals remains largely
within each layer \cite{Yao_PRL_2023}. This makes bilayer nickelates
a promising platform for exploring unconventional superconductivity and spin\textendash layer\textendash orbital
entanglement phenomena beyond the paradigms established in cuprates \cite{Liao_PRB_2023,QuSu_arXiv_2023, WangHu_arXiv_2024,
  HeierSavrasov_PRB_2024, ZhanHu_arXiv_2024,
  ChangLi_arXiv_2023,
  JiangZhang_CPL_2024, HuangZhou_PRB_2023,
  XueWang_CPL_2024, ChenLi_PRB_2024, KanekoKuroki_PRB_2024,
  SakakibaraKuroki_PRL_2024,
  JiangKu_PRL_2024, LiuChen_arXiv_2023, YangZhang_arXiv_2024,
  YangZhang_arXiv_2023, ZhangWeng_PRL_2024, LuYou_arXiv_2023, FanXiang_PRB_2024,
  ZhengWu_arXiv_2023, SchlomerBohrdt_arXiv_2023, BotzelEremin_arXiv_2024_1,
  OhZhang_arXiv_2024,hu2025film,yang2025film1,
  QuSu_PRL_2024, PanWu_arXiv_2023, WangYang_arXiv_2024, QinYang_PRB_2023,
  LuoYao_npjQM_2024, YangZhang_PRB_2023, LuWu_PRL_2024, LuWu_PRB_2024,
  KakoiKuroki_PRB_2024, MaWu_arXiv_2024, YangWang_PRB_2023, ZhangDagotto_NC_2024,ji2025strong,kaneko2025t}.

More generally, in
transition-metal systems with active $e_g$ orbitals,
the bilayer geometry naturally leads to a phenomenon of strongly
orbital-selective
interlayer hybridization. Due to the directional nature of $d$-orbitals,
electronic coupling along the stacking direction is highly anisotropic:
the orbitals extending vertically (
{\it e.g.} the $d_{z^{2}}$ orbitals) feel a robust
interlayer overlap, whereas the planar orbitals (
{\it e.g.} the $d_{x^{2}-y^{2}}$ orbitals)
between upper and lower layers remain effectively decoupled
\cite{Liao_PRB_2023,Yao_PRL_2023,liao2412}. This
intrinsic energy hierarchy
not only reshapes the band structure,
but also acts as an orbital
filter that can dynamically quench specific orbital sectors at low
energies. As a result, the system enters a unique regime where the
surviving layer degree of freedom intertwines with spin, setting the
stage for exotic composite entanglement.

In this work, we capitalize on this geometric mechanism to derive
a  low-energy theory for a quarter-hole-filled two-orbital bilayer Hubbard
model.
By explicitly treating the strong interlayer hybridization of $d_{z^{2}}$
orbitals within a molecular-orbital basis, we project out the high-energy
spin singlet excitations and construct an effective anisotropic
Kugel-Khomskii Hamiltonian
describing the coupled dynamics of electron spin and emergent layer
pseudospin degrees of freedom associated with the $d_{x^{2}-y^{2}}$
orbitals in the upper and lower layers. Through
a combination of Weiss mean-field theory and generalized flavor-wave
theory, we uncover a rich phase diagram. Most notably, we identify
a novel spin-layer-entangled (SLE) phase in the strong spin-layer
coupling regime. In this state, conventional long-range magnetic and
layer orders are simultaneously melted, giving way to a  hidden composite
order characterized by maximal local entanglement between spin and
layer sectors. We further show that this phase arises from the spontaneous
breaking of an emergent $O(4)$ symmetry, manifesting in a unique
excitation spectrum with three gapless Goldstone modes.

\section{Model and hamiltonian}

\subsection{Bilayer two-orbital Hubbard model}

We consider a bilayer two-orbital Hubbard model
that captures
the essential low-energy physics of transition-metal bilayer systems
with strong orbital-selective interlayer hybridization. Here we take the
two-orbital bilayer Hubbard model for the nickelate superconductor
La$_{3}$Ni$_{2}$O$_{7}$ as a representative example \cite{
  KanekoKuroki_PRB_2024,SakakibaraKuroki_PRL_2024}.

In La$_{3}$Ni$_{2}$O$_{7}$, the $t_{2g}$ orbitals are fully occupied
and each layer hosts two $e_g$ orbitals, $d_{x^{2}-y^{2}}$ and $d_{z^{2}}$,
with the average electron filling corresponding to three electrons
per bilayer rung, {\it e.g.}, a quarter-hole-filled configuration \cite{Sun_Nature_2023,Yao_PRL_2023}. The kinetic
properties of these $e_{g}$ electrons are strictly governed by the
highly spatially anisotropic nature of the atomic orbitals: The wave function of $d_{x^{2}-y^{2}}$
orbitals is extended within the crystal plane, with lobes pointing
towards the in-plane oxygen ligands. This geometry facilitates strong
intralayer hybridization. Meanwhile, the interlayer overlap between $d_{x^{2}-y^{2}}$
orbitals is negligible due to the lack of vertical extension of the wave
function. In stark contrast,
the $d_{z^{2}}$ orbitals feature lobe-shaped electron densities
elongated along
the $c$-axis, enabling a distinct
interlayer hopping channel that
is characteristic of
$d_{z^{2}}$
symmetry \cite{liao2412,cui2024strain,duan2025orbital}.

\begin{figure}
  \includegraphics[width=1\linewidth]{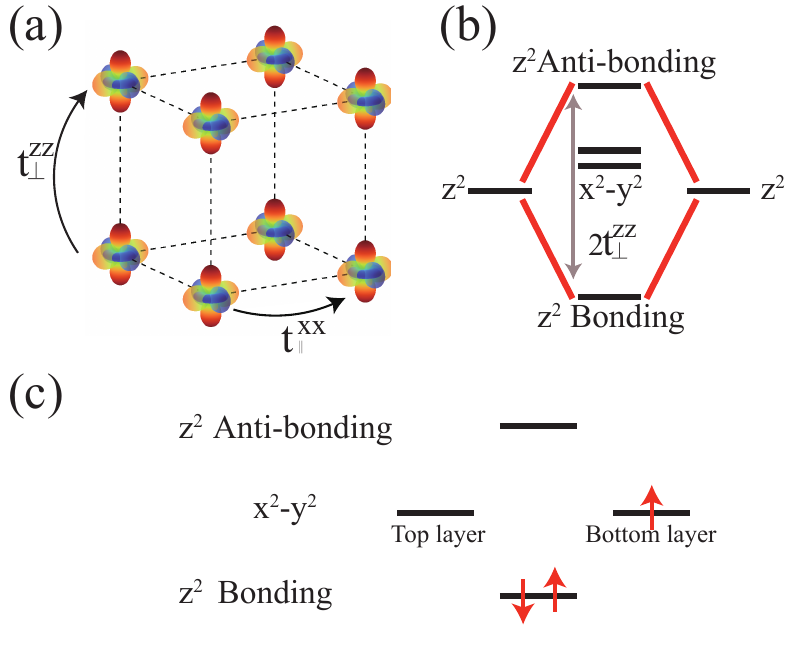}

  \caption{(a) Schematic illustration of the minimal bilayer two-orbital
  tight-binding
  model where $t^{zz}_{\perp}$ and $t^{xx}_{\parallel}$ denote the
  orbital dependent hopping parameters.
  (b) Sketch of the crystal splitting of $e_g$ orbitals and the formation of the
  bonding-antibonding molecular orbital (MO) states
  between Ni $z^2$ orbitals in the top and bottom layers. (c)
  One representative
  ground-state configuration
  where the bonding $d_{z^{2}}$ orbital is doubly occupied, while
  the remaining electron resides in one of the $d_{x^{2}-y^{2}}$ orbitals.
  The other configurations
  can be obtained
  by reversing the spin direction or the layer occupation from those
  presented in (c).
  }

  \label{fig1}
\end{figure}

Based on the orbital-selective physics, we start from the
bilayer
within the two
$e_g$
orbital sector.
The Hamiltonian reads
\begin{equation}
  H=H_{\rm{TB}}+H_{\rm{int}}. \label{eq:0}
\end{equation}
Here, $H_{\rm{TB}}$ is a minimal tight-binding
Hamiltonian (as sketched in Fig.~\ref{fig1}(a)):
\begin{equation}
  H_{\rm{TB}}=\sum_{ij\sigma\eta} t_{\parallel}^{xx}d_{ix\sigma}^{\eta\dagger}
  d_{jx\sigma}^{\eta} +\sum_{i\sigma}t_{\perp}^{zz}
  d_{iz\sigma}^{T\dagger}d_{iz\sigma}^{B},\label{eq:1}
\end{equation}
where $d_{i\alpha\sigma}^{\eta\dagger}\ (d_{i\alpha\sigma}^{\eta})$
creates (annihilates) an electron in orbital $\alpha$ ($\alpha=x,z$
denotes the two $e_g$ orbitals, $d_{x^{2}-y^{2}}$ and $d_{z^{2}}$,
respectively) with spin $\sigma$ at site $i$ of layer $\eta$ ($\eta=$T,
B denotes the top and bottom layer, respectively). The hopping amplitude
$t_{\parallel}^{xx}$ is the nearest-neighbor intralayer hopping of
the $d_{x^{2}-y^{2}}$ orbitals, while $t_{\perp}^{zz}$ captures
the strong interlayer hybridization between vertically aligned $d_{z^{2}}$
orbitals. Direct interlayer hopping of the $d_{x^{2}-y^{2}}$ orbitals
is neglected due to their planar orbital character. The presence of
strong interlayer hopping between the $d_{z^{2}}$ orbitals $t_{\perp}$
has a significant effect on the single-site spectrum, and
will be discussed in the next subsection.

The on-site term $H_{\rm{int}}$ consists of two contributions, the crystal
field splitting and the local electron-electron interactions, which
are taken in the Kanamori form \cite{georges2013strong}:

\begin{align}
  H_{\rm{int}} &
  =\sum_{i\alpha\eta}\epsilon_{\alpha} n_{i\alpha}^{\eta}
  +U\sum_{i,\alpha,\eta}n_{i\alpha\uparrow}^{\eta}
  n_{i\alpha\downarrow}^{\eta}\nonumber                        \\
               & +\sum_{i,\alpha<\beta,\sigma,\eta} \{U^\prime
  n_{i\alpha\sigma}^{\eta}
  n_{i\beta\bar{\sigma}}^{\eta}
  +(U^\prime-J_{\rm{H}})n_{i\alpha\sigma}^{\eta}
  n_{i\beta\sigma}^{\eta}\label{eq:2}                          \\
               & -J_{\rm{H}}(d_{i\alpha\sigma}^{\eta\dagger}
  d_{i\alpha\bar{\sigma}}^{\eta}
  d_{i\beta\bar{\sigma}}^{\eta\dagger} d_{i\beta\sigma}^{\eta}
  +d_{i\alpha\sigma}^{\eta\dagger} d_{i\alpha\bar{\sigma}}^{\eta\dagger}
  d_{i\beta\sigma}^{\eta} d_{i\beta\bar{\sigma}}^{\eta})\}\nonumber
\end{align}
where $n_{i\alpha\sigma}^{\eta}=d_{i\alpha\sigma}^{\eta\dagger}d_{i\alpha\sigma}^{\eta}$.
Here $U$, $U^\prime$ and $J_{\rm{H}}$, represent the intra- and inter-orbital
Coulomb repulsion and the Hund's coupling, respectively, satisfying
$U^\prime=U-2J_{\rm{H}}$.
We assume a crystal field splitting $\epsilon_{x}>\epsilon_{z}$,
consistent with first-principles results \cite{Yao_PRL_2023} for bilayer nickelates (Fig.
\ref{fig1} (b)).

\subsection{Molecular Orbital Basis in the Strong Coupling Limit}

A unique feature of this La$_{3}$Ni$_{2}$O$_{7}$ bilayer system
is the significant orbital-selective interlayer hybridization along
with fractional electron occupation per Ni ion. While the interlayer
coupling between $d_{x^{2}-y^{2}}$ orbitals is negligible,  the strong
interlayer hopping  of the $d_{z^{2}}$ orbitals $t_{\perp}$ leads
to the formation of bonding-antibonding molecular orbitals (MOs), illustrated in Fig. \ref{fig1} (b).
To capture this effect, we work with the MO basis of $d_{z^{2}}$
orbitals

\begin{equation}
  d_{iz\sigma}^{b(a)}=\frac{1}{\sqrt{2}}(d_{iz\sigma}^{T}\pm
  d_{i+\delta_{z}z\sigma}^{B}),
\end{equation}
where the index $b$($a$) corresponds to the bonding (antibonding)
MO. In the MO basis, the on-site energy of $d_{z^{2}}$ orbital is
renormalized to $\epsilon_{z}\mp t_{\perp}^{zz}$. This energy separation
is the crucial mechanism that stabilizes the low-energy manifold.

\begin{figure}
  \includegraphics[width=1\linewidth]{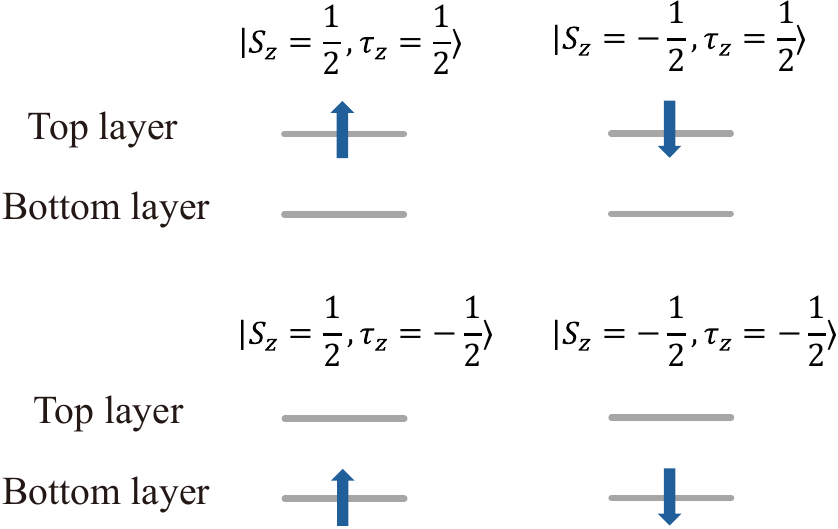}
  \caption{Four
  degenerate ground-state configurations $|S^{z},\ \tau^{z}\rangle$ labeled
  with the spin and layer quantum numbers
  $S^{z}=\pm\frac{1}{2}$ and $\tau^{z}=\pm\frac{1}{2}$ in the
  $d_{x^{2}-y^{2}}$ orbital subspace.
  }
  \label{fig2}
\end{figure}

The Hamiltonians of Eq. \eqref{eq:1} and Eq. \eqref{eq:2} are recast
in the MO basis as $H=H_{\rm{TB}}^{\rm{MO}}+H_{\rm{int}}^{\rm{MO}}$, where the
interaction
terms are transformed accordingly. We focus on the quarter-hole filling
regime, corresponding to a total occupancy of $n=3$ electrons per
rung.
Theoretical calculations consistently suggest that the ground-state
configuration relevant to La$_{3}$Ni$_{2}$O$_{7}$
is a low-spin state \cite{liao2412,
  TianLu_PRB_2024,Yao_PRL_2023,Sun_Nature_2023} where the bonding $d_{z^{2}}$
orbital
is doubly occupied by a spin-singlet and the degenerate $d_{x^{2}-y^{2}}$
orbitals
are quarterly filled, as
illustrated in Fig. \ref{fig1}(c). Note that this ground-state configuration
is four-fold degenerate, reflecting the fact that the remaining electron
can occupy the $d_{x^{2}-y^{2}}$ orbital on either the top or bottom
layer with either spin orientation. To capture the layer degree of
freedom of the $d_{x^{2}-y^{2}}$ orbitals, we introduce a layer pseudospin
operator $\tau$, where $\tau^{z}=\pm\frac{1}{2}$ characterizes the
occupation on the top and the bottom layers, respectively. The four-fold
low-spin configurations are therefore labelled by the $|S^{z},\ \tau^{z}\rangle$,
where $S^{z}=\pm\frac{1}{2}$ and $\tau^{z}=\pm\frac{1}{2}$ denote
the total electron spin and layer pseudospin, respectively. The schematic illustration is shown in Fig. \ref{fig2}.

\subsection{Schrieffer-Wolff transformation and Second perturbation}

Given the strong coupling limit where the in-plane hopping amplitude
is significantly smaller than the interaction scale ({\it e.g.}
$t_{\parallel}^{xx}\ll U$),
we treat the in-plane kinetic term $H^\prime$ as a perturbation to the
local Hamiltonian $H_{0}$. In this regime, the ground state manifold
is determined primarily by local interactions, and the effects of
itinerancy enter only through higher order virtual processes, justifying
a controlled strong-coupling approach. To capture the resulting low-energy
physics within the degenerate ground-state manifold, we derive an
effective Hamiltonian via a Schrieffer\textendash Wolff transformation,
$H_{\text{eff}}=e^{S}He^{-S}$, which eliminates high-energy charge
excitations and projects the dynamics onto the low-energy subspace
$\mathcal{S}$ \cite{schrieffer1966relation}.

The leading contribution to the effective interaction arises from
second-order virtual hopping processes \cite{anderson1950antiferromagnetism},
$H_{\text{eff}}\approx\frac{1}{2}[S,H^\prime]$.
Starting from a product of local ground-state configurations
$|S^{z},\tau^{z}\rangle_{i}\otimes|S^{z},\tau^{z}\rangle_{j}$,
the hopping term $t_{\parallel}^{xx}$ transfers an electron in the
$d_{x^{2}-y^{2}}$ orbital
to a neighboring site, generating a high-energy intermediate
state. The large energy cost of this intermediate state originates
not only from the Coulomb repulsion $U$ but also from the suppression
of Hund's coupling: Specifically, the rigid spin singlet formed by
the electrons in the bonding $d_{z^{2}}$ orbital prevents the itinerant
$d_{x^{2}-y^{2}}$ electron from aligning ferromagnetically with the
$d_{z^{2}}$ electrons, thereby suppressing the Hund's energy gain.

\begin{figure}
  \includegraphics[width=1\linewidth]{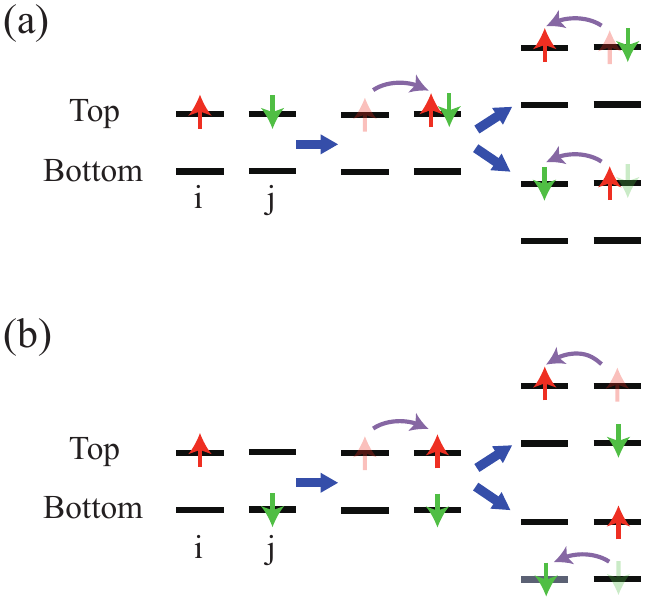}

  \caption{Schematic illustration of the major virtual hopping processes in the
    second-order perturbation expansion. For clarity, the bonding $d_{z^{2}}$
    orbitals are omitted as they remain fully occupied throughout the
    process. (a) Type-I: Processes involving electrons in the same layer.
    (b) Type-II: Processes involving electrons in different layers.}

  \label{fig3}
\end{figure}

We classify these virtual fluctuations into two channels based on
the layer configuration as shown in Fig. \ref{fig3}. In Type-I processes
($\tau_{i}^{z}=\tau_{j}^{z}$), electrons reside in the same layer.
The virtual hopping creates a double occupancy that strictly conserves
the layer index $\tau_{z}$ while mediating spin exchange. In Type-II
processes ($\tau_{i}^{z}\neq\tau_{j}^{z}$), electrons reside in opposite
layers. Here, the return hop allows for mixing between different basis
states, mediating coupled fluctuations of both spin and layer pseudospin.

Integrating these processes yields an effective anisotropic Kugel-Khomskii type
Hamiltonian:
\begin{align}
  H_{\text{eff}} &
  =\sum_{ij}J_{s}\boldsymbol{\hat{S}_{i}}\cdot\boldsymbol{\hat{S}_{j}}
  +\sum_{\alpha}K_{s}^{\alpha\alpha}(\frac{\text{1}}{4}
  -\boldsymbol{\hat{S_{i}}}\cdot\boldsymbol{\hat{S_{j}}})
  \hat{\tau}_{i}^{\alpha}\hat{\tau}_{j}^{\alpha}\nonumber         \\
                 & +K_{t}^{\alpha\alpha}(\boldsymbol{\hat{S}_{i}}
  \cdot\boldsymbol{\hat{S}_{j}}
  +\frac{\text{3}}{4})\hat{\tau}_{i}^{\alpha}
  \hat{\tau}_{j}^{\alpha}.\label{eq:H_eff}
\end{align}
Due to the complex multiplet structure of the intermediate states,
analytical expressions for the exchange constants are unwieldy. Instead,
we can determine the relationships between effective parameters $\{K_{s}^{\alpha\alpha},K_{t}^{\alpha\alpha},J_{s}\}$
and $U$, $J_{\rm{H}}$, $t_{\perp}^{zz}$, $\epsilon_\alpha$ by numerically
diagonalizing
the
associated two-site Hubbard Hamiltonian.

It is important to note that $\tau^{z}$ is not a strictly conserved
quantum number when inter-orbital fluctuations are taken into account. As a result, the system
does not preserve a $U(1)$ rotational symmetry in the
layer sector. Consequently, the transverse exchange couplings are
anisotropic, i.e., $K_{s/t}^{xx}\neq K_{s/t}^{yy}$. Nevertheless,
since the $\tau^{z}$ violations are suppressed by the small amplitudes
of the mixed layer components, this anisotropy is quantitatively negligible.
That is, the system retains an approximate $U(1)$ symmetry with
$K^{xx}\approx K^{yy}$.
In summary, the effective model, in the $K^{xx}=K^{yy}$ limit, has an $O(3)_\text{spin}\times O(2)_\text{layer}$ symmetry, where the layer  $O(2)_\text{layer}$ symmetry comes from the combination of two distinct symmetries:
the continuous $U(1)$ symmetry in the transverse
channel $\mathcal G(\theta)\equiv\exp[i\sum_i \hat\tau^z_i \theta]$, and a discrete $\mathbb{Z}_2$ symmetry associated with the exchange between top and bottom layers, i.e., $\mathcal I\equiv \prod_i \hat\tau^x_i$.
The spontaneous breaking of these symmetries gives rise to different ordered
phases, as will be discussed in the next section.

\section{Ground-state phase diagram of the effective Kugel-Khomskii model}

\subsection{Weiss mean-field theory}

In this section, we investigate the ground-state phase diagram of
the effective anisotropic Kugel-Khomskii model of Eq. \eqref{eq:H_eff}
within the Weiss mean-field
theory. This formalism is equivalent to adopting a
site factorized trial
wavefunction
\begin{equation}
  |\Psi\rangle=\otimes_{i}|\boldsymbol{d}\rangle_{i}.
\end{equation}
Here $|\boldsymbol{d}\rangle_{i}$ is the local wave function at site $i$
and can be expressed as a coherent superposition of the basis states
$|S^{z},\tau^{z}\rangle_{i}$,
\begin{align}
  |\boldsymbol{d}\rangle_{i} & =\sum_{S^{z},\tau^{z}=
    \pm\frac{1}{2}}d_{i,S^{z},\tau^{z}}|S^{z},\tau^{z}\rangle_{i},
\end{align}
where $d_{i,S^{z},\tau^{z}}$ are complex variational coefficients.
The ground state can then be determined by variationally minimizing the
energy $E=\langle \Psi|H|\Psi\rangle/\langle \Psi|\Psi \rangle$.
While this trial wave function does not capture the quantum
entanglement between different sites, this treatment retains the on-site
entanglement between spin and layer pseudospin degrees of freedom.
As a result, it is capable of describing exotic spin\textendash layer\textendash entangled
states, as will be discussed later.

To explore the generic features of the model without being restricted
to a specific material La$_{3}$Ni$_{2}$O$_{7}$, we treat the interaction
strengths in Eq.\eqref{eq:H_eff} as independent parameters. Owing
to the high dimensionality of the parameter space, our investigation
focuses on a representative cross-section that captures the essential
physics.

\begin{figure}
  \includegraphics[width=1\linewidth]{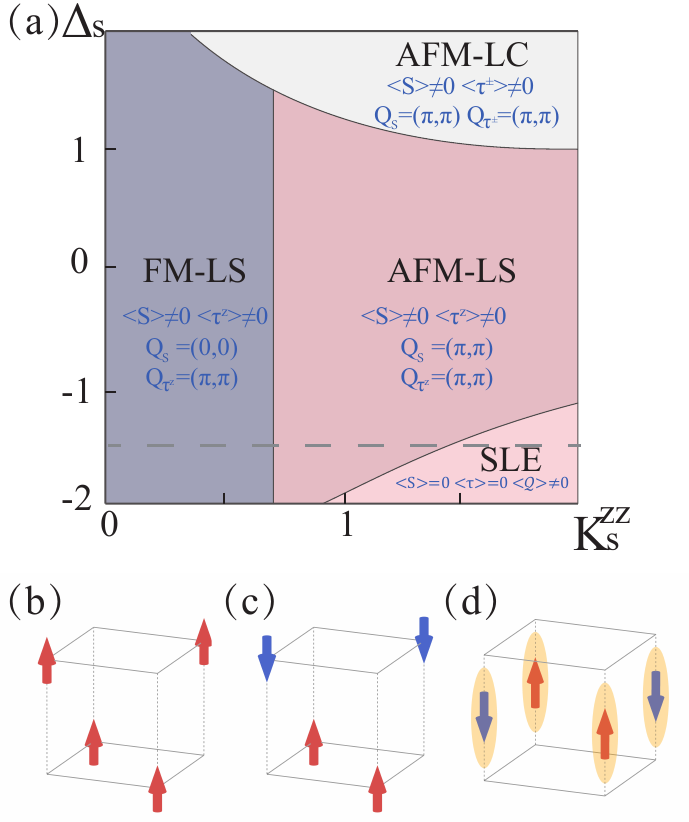}
  \caption{(a) Ground-state phase diagram of the effective Kugel-Khomskii
  model in Eq.~\ref{eq:H_eff}. The axes represent
  $K_{s}^{zz}$ and $\Delta_{s}=K_{s}^{xx}/K_{s}^{zz}$ (assuming $K_{s}^{xx}=K_{s}^{yy}$).
  The other parameters are fixed to $J_{s}=0.2$, $K_{t}^{zz}=1.5$,
  and $K_{t}^{xx}=K_{t}^{yy}=1.2$. Four phases are stabilized,
  which are denoted as
  FM-LS (spin-ferromagnetic and layer-staggered), AFM-LS (spin-antiferromagnetic
  and layer-staggered), AFM-LC (spin-antiferromagnetic and interlayer
  coherent), and SLE (spin-layer-entangled), respectively. Here,
  $\mathbf{Q}_S$ and
  $\mathbf{Q}_{\tau^\alpha}$ denote the ordering momenta where
  the corresponding spin and layer structure
  factors exhibit Bragg peaks, respectively.
  (b) Configuration of the FM-LS phase,
  where a spin-ferromagnetic state coexists with spatially staggered layer
  occupation, manifesting as a checkerboard charge pattern. (c) Configuration
  of the AFM\textendash LS
  phase, where the spin-antiferromagnetic state
  retains
  a staggered layer occupation. (d) AFM\textendash LC phase, where the
  spin-antiferromagnetic state is accompanied by
  a layer-coherent order. Here, the pseudospins lie in the $xy$-plane,
  representing spontaneous quantum coherence between the top and bottom
  layers.}
  \label{fig4}
\end{figure}

We fix the exchange scales in the pure spin and spin-layer triplet
channels to $J_{s}=0.2$, $K_{t}^{zz}=1.5$, and $K_{t}^{xx}=K_{t}^{yy}=1.2$,
and then systematically investigate the influence of the coupling
$K_{s}^{\alpha\alpha}$ on the system's ground state. For simplicity,
we assume $K_{s}^{xx}=K_{s}^{yy}=\Delta_{s}K_{s}^{zz}$. The phase
diagram is shown in Fig. \ref{fig4}, which exhibits four distinct phases:
(1) spin-ferromagnetic and layer staggered (FM-LS), (2) spin-antiferromagnetic
and layer staggered (AFM-LS), (3) spin-antiferromagnetic and interlayer
coherent (AFM-LC), and (4) spin-layer-entangled (SLE) phase. Configurations of
the first three phases are illustrated in panels (b)-(d) of Fig.~\ref{fig4}.

\subsection{Layer staggered phases}

We first consider the regime dominated by the Ising pseudospin anisotropy where $|\Delta_s|$ is small.
Throughout this regime, the system spontaneously breaks the pseudospin
$\mathbb{Z}_2$ symmetry $\langle \hat\tau^z \rangle\neq 0$, forming
layer-staggered
electron occupation where electrons alternately populate the top and
bottom layers on adjacent sites, i.e., $\langle\hat{\tau}_{i}^{z}\hat{\tau}_{j}^{z}\rangle=-\frac{1}{4}$.
This pattern manifests as a checkerboard charge order in the two layers.
In addition, the spin sectors can exhibit either ferromagnetic or antiferromagnetic orders,
depending on the sign of the coupling strength $K_{s}^{zz}$, as shown in
Fig. \ref{fig4} (b) and (c).

Within the layer-staggered phases, the ground state energy can be written as:

\begin{equation}
  E_{\text{LS}}=\frac{4J_{s}-K_{t}^{zz}+K_{s}^{zz}}{4}\langle\boldsymbol{\hat{S}_{i}}\cdot\boldsymbol{\hat{S}_{j}}\rangle-
  (\frac{K_{s}^{zz}}{16}+\frac{3K_{t}^{zz}}{16}).
\end{equation}

A transition from the FM to the AFM spin configuration is driven by the competition between the effective exchange interactions
$K_s^{zz}$ and $K_t^{zz}$.
In particular, the AFM order becomes energetically favorable when $K_{s}^{zz}+4J_{s}>K_{t}^{zz}$.

\subsection{Inter-layer coherent phase}

In another limit where $\Delta_{s}$ is large and the pseudospin exhibits XY anisotropy,
the system tends to break the pseudospin U(1) symmetry and establish off-diagonal correlations in the pseudospin sector: $\langle\hat{\tau}^{\pm}\rangle \neq 0$.
This signals the emergence of spontaneous inter-layer coherence,
in which electrons form coherent quantum superpositions of occupying
the top and bottom layers, see Fig. \ref{fig4} (d).
In the mean time, spins are ordered antiferromagnetically to synergistically
minimize the total energy.
The ground state energy per bond is:

\begin{equation}
  E_{\text{c}}=-\frac{J_{s}}{4}-\frac{1}{8}(K_{t}^{xx}+K_{s}^{xx}).
\end{equation}
Physically, the vanishing of $\langle\hat{\tau}^{z}\rangle$ implies
the melting of the spatially staggered layer occupation. The emergence
of in-plane layer pseudospin order corresponds to a state with spontaneous
inter-layer coherence, where electrons form a quantum superposition
between the top and bottom layers with a specific relative phase.
In contrast to the layer staggered phase, this coherent state preserves
layer symmetry and therefore does not induce any electric polarization or
static charge order in real space.
Moreover, since the inter-layer coherent order resides entirely in the
off-diagonal pseudospin channel, it does not gap out the charge sector.
As a result, no charge excitation gap associated with layer polarization
is generated in this phase,
as shown in the next section.

\subsection{Spin-Layer-Entangled}

\begin{figure}

  \includegraphics[width=1\linewidth]{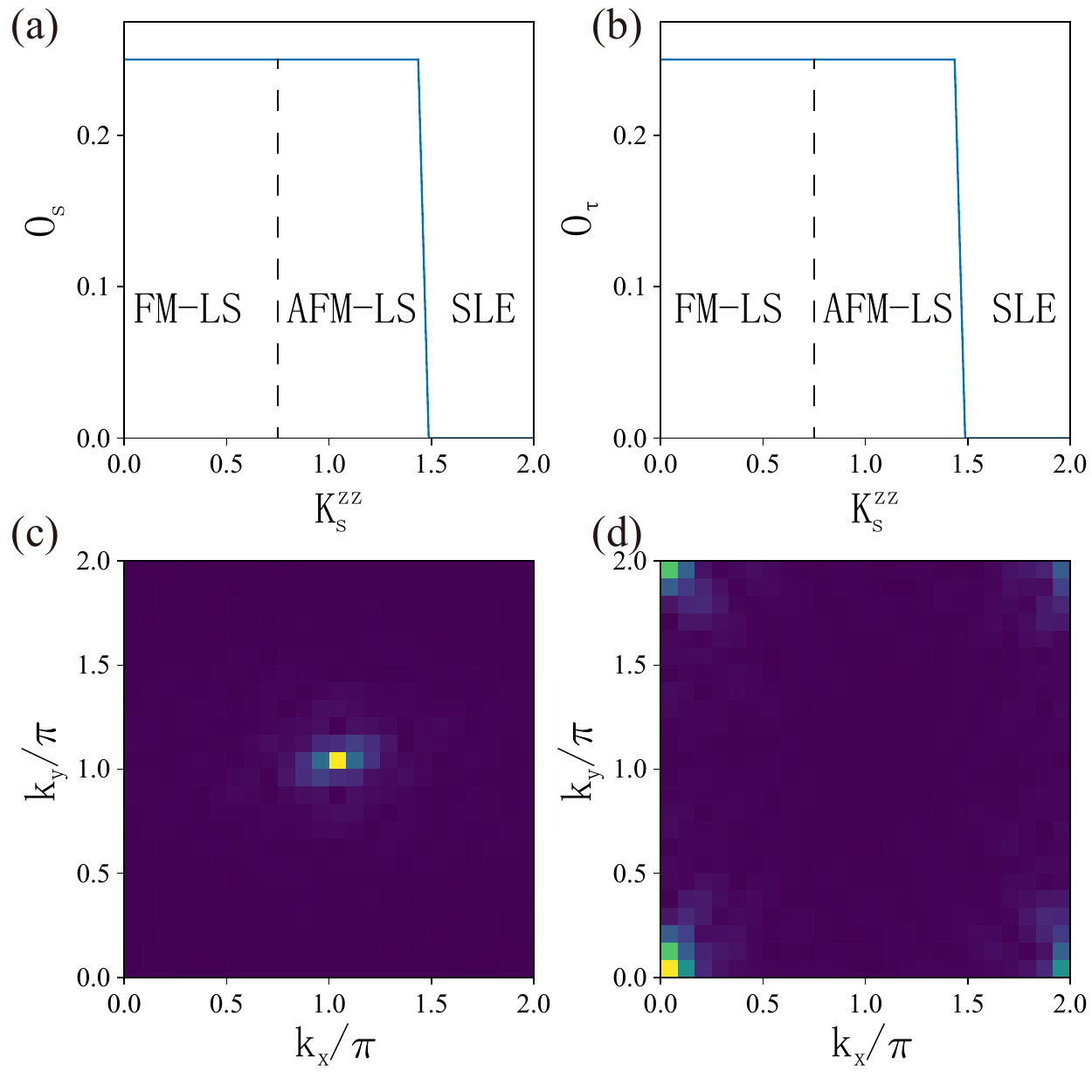}\caption{(a), (b) Averaged local moments in the spin and pseudospin sectors along the dashed line in Fig. \ref{fig4} (a),
  defined as $\langle O_{s}\rangle=\frac{1}{L^{2}}\sum_{i,\alpha}\langle S_{i}^{\alpha}\rangle^{2}$
  and $\langle O_{\tau}\rangle=\frac{1}{L^{2}}\sum_{i,\alpha}\langle\Gamma_{i}^{\alpha}\rangle^{2}$,
  respectively. (c), (d) Momentum-space distribution of spin-layer correlations
  $S_{\mu}(\mathbf{k})=\frac{1}{L^{2}}\sum_{i}e^{i\mathbf{k}\cdot\mathbf{r}_{i}}\sum_{\alpha,\beta}\mathcal{Q}_{i}^{\beta\alpha}$,
  where $\beta\in\{x,y,z\}$. (c) shows the transverse component ($\mu=\perp$)
  summing over $\alpha\in\{x,y\}$, and (d) shows the longitudinal component
  ($\mu=\parallel$) with $\alpha=z$.}

  \label{fig5}

\end{figure}

Interestingly, upon decreasing the anisotropy $\Delta_{s}$, the system enters an intermediate regime
where the spin and layer degrees of freedom become strongly entangled.
This phase breaks both the inversion $\mathcal I$ and time reversal symmetries. However,
such symmetry breaking pattern cannot be characterized by the spins or
pseudospins order alone: In this phase,
$\langle\hat{\boldsymbol{S}}_{i}\rangle=0$
and $\langle\hat{\boldsymbol{\tau}}_{i}\rangle=0$.
Instead, it is necessary
to introduce
a composite spin\textendash layer order parameter
$\mathcal{\hat{Q}}_{i}^{\beta\alpha}\equiv
  \hat{S}_{i}^{\beta}\hat{\tau}_{i}^{\alpha}$,
with $\alpha,\beta\in\{x,y,z\}$, to describe the symmetry breaking of this phase.
The composite spin\textendash layer order can be well understood by rewriting
the Hamiltonian Eq. (\ref{eq:H_eff}) in terms of these operators:
\begin{align*}
  H_{\text{eff}} & =\sum_{ij}J_{s}\boldsymbol{\hat{S}_{i}}\cdot\boldsymbol{\hat{S}_{j}}+\sum_{\alpha}(\frac{\text{\ensuremath{K_{s}^{\alpha\alpha}}}}{4}+\frac{\text{3\ensuremath{K_{t}^{\alpha\alpha}}}}{4})\hat{\tau}_{i}^{\alpha}\hat{\tau}_{j}^{\alpha}+ \\
                 & \sum_{\alpha\beta}(\text{\ensuremath{K_{t}^{\alpha\alpha}}}-K_{s}^{\alpha\alpha})\hat{\mathcal{Q}}_{i}^{\beta\alpha}\hat{\mathcal{Q}}_{j}^{\beta\alpha}.
\end{align*}
In
the SLE regime of the phase diagram,
the $\hat{\mathcal{Q}}_{i}\hat{\mathcal{Q}}_{j}$ term dominates. As a result,
a spin-layer-entangled order
with $\langle\mathcal{\hat{Q}}_{i}^{\beta\alpha}\rangle\neq 0$ is established in
the ground state. Meanwhile, the spin and pseudospin are disordered,
$\langle\hat{\boldsymbol{S}}_{i}\rangle=0$
and $\langle\hat{\boldsymbol{\tau}}_{i}\rangle=0$, as depicted in
Fig.~\ref{fig5} (a, b).
Despite this, all nine components of $\mathcal{Q}_{i}^{\beta\alpha}$
are generally non-zero.
The relation that $\langle\mathcal{\hat{Q}}_{i}^{\beta\alpha}\rangle=\langle\hat{S}_{i}^{\beta}\hat{\tau}_{i}^{\alpha}\rangle\neq\langle\hat{S}_{i}^{\beta}\rangle\langle\hat{\tau}_{i}^{\alpha}\rangle=0$
signals the emergence of a nontrivial composite ordered phase with  a maximal spin-layer entanglement.
In fact, we find that the on-site ground-state wavefunction obtained by the Weiss mean-field theory satisfies the following form

\begin{align}
  |\boldsymbol{d}\rangle_{i} & = a|+\frac{1}{2},+\frac{1}{2}\rangle_{i}+a^*|-\frac{1}{2},-\frac{1}{2}\rangle_{i} \nonumber \\
                             & + i b|+\frac{1}{2},-\frac{1}{2}\rangle_{i}+i b^*|-\frac{1}{2},+\frac{1}{2}\rangle_{i}
\end{align}
for the A sublattice and
\begin{align}
  |\boldsymbol{d}\rangle_{i} & = ia|+\frac{1}{2},+\frac{1}{2}\rangle_{i}-ia^*|-\frac{1}{2},-\frac{1}{2}\rangle_{i} \nonumber \\
                             & + b|+\frac{1}{2},-\frac{1}{2}\rangle_{i}-i (ib)^*|-\frac{1}{2},+\frac{1}{2}\rangle_{i}
\end{align}
for the B sublattice,
where $a$ and $b$ are complex numbers satisfying $|a|^2+|b|^2=\frac{1}{2}$. This indicates that one can alternatively define order parameter for the SLE phase as a four-component real scalar $\boldsymbol\phi\equiv(\Re a, \Im a, \Re b, \Im b)^T$.
Hence the order parameter $\boldsymbol\phi$ lives in emergent $S^3/\mathbb{Z}_2$ manifold, as they can be arbitrary chosen on a four-sphere with radius $\sqrt{1/2}$, and that $\boldsymbol\phi$ and $-\boldsymbol\phi$ corresponds to the same state.
Also, the mean-field ground state possesses an emergent $O(4)$ symmetry, although the Hamiltonian only has a smaller $O(3)_\text{spin}\times O(2)_\text{layer}$ symmetry.

From the mean-field wave function, we observe that the composite order exhibits a characteristic spatial texture depending on the layer pseudospin index $\alpha$.
In particular, the transverse components display an antiferromagnetic pattern (Fig. \ref{fig5}(c)), satisfying $\mathcal{Q}_{i}^{\beta\alpha}=-\mathcal{Q}_{i+\delta}^{\beta\alpha}$
for $\alpha\in\{x,y\}$, whereas the longitudinal component shows
a ferromagnetic distribution, $\mathcal{Q}_{i}^{\beta z}=\mathcal{Q}_{i+\delta}^{\beta z}$,
as illustrated in Fig. \ref{fig5}(d).

Despite strong fluctuations in the individual spin and pseudospin degreees of freedom, we find that their relative orientation is rigidly locked. This property can be made clear by introducing a composite pseudospin operator
\begin{equation}
  \boldsymbol{\mathcal{J}}_{i}=\hat{\boldsymbol{S}}_{i}-4\mathcal{Q}_{i} {\hat{\boldsymbol\tau}}_{i}.
\end{equation}
One can verify that $\mathcal{J}_{i}$ satisfies the angular momentum algebra $[\mathcal{J}_{i}^\alpha,\mathcal{J}_{i}^\beta]=i\epsilon^{\alpha\beta\gamma}\mathcal{J}_{i}^\gamma$.
Further, it is straightforward to check that $\boldsymbol{\mathcal{J}}_{i}^2|\boldsymbol{d}\rangle_{i}=0$, which indicates that the fluctuating spin ${\boldsymbol{S}}_{i}$ and the transformed pseudospin degrees of freedom $-4\mathcal{Q}_{i} {\hat{\boldsymbol\tau}}_{i}$ are locked antiferromagnetically.
The system thus suppresses relative spin\textendash layer fluctuations to minimize the strong coupling energy, giving rise to a hidden composite order that is not detectable by conventional probes of dipolar magnetism.

\section{Excitation spectrum of the effective Kugel-Khomskii model}

\begin{figure}[H]
  \includegraphics[width=1\linewidth]{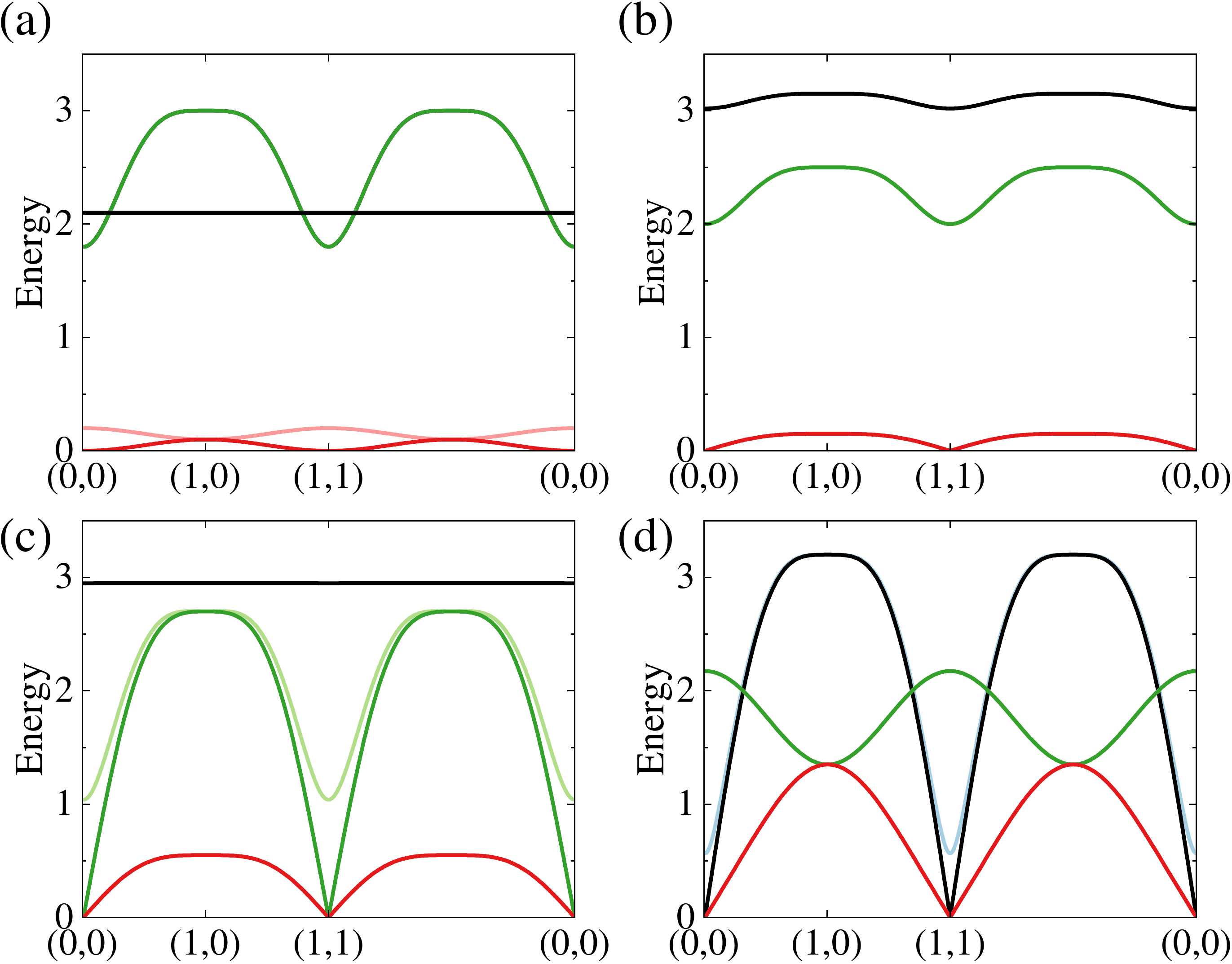}

  \caption{The excitation dispersion along the high symmetry lines in different
    phases. (a) FM-LS phase with $K_{s}^{zz}=0.5$, $\Delta_{s}=0.3$;
    (b) AFM-LS phase with $K_{s}^{zz}=1.0$, $\Delta_{s}=0.3$; (c) AFM-LC
    phase with $K_{s}^{zz}=1.0$, $\Delta_{s}=1.5$; (d) SLE phase with
    $K_{s}^{zz}=1.0$, $\Delta_{s}=-2$. Here, the red and green lines denote the spin and layer excitations, respectively. The black lines relate to the composite spin-layer excitations.}

  \label{fig6}
\end{figure}

To better understand the nature of the different phases in the phase diagram,
here we discuss the excitation spectrum within each phase presented in the
ground-state phase diagram of Fig.~\ref{fig4}(a).
We employ a generalized spin-wave theory to describe quasiparticle
excitations of this model \cite{dahlbom2025sunny}.
Compared to ordinary spin-wave theories where spins and pseudospins are
independently treated via the Holstein-Primakoff transformation in each sector,
the generalized spin-wave theory here correctly captures the on-site
spin-pseudospin entanglement.
The details of the formalism are presented in Appendix B. The calculated excitation dispersions for the four phases are presented in Fig. \ref{fig6}.

For both the FM-LS and AFM-LS phases, the excitation spectra exhibit clear separation of energy scales, indicating different origins of excitations.
The low-energy gapless excitations (red line in Fig. \ref{fig6}(a, b)) are assigned as spin excitations that arise from the spontaneous breaking of the continuous $O(3)$ spin-rotational symmetry.
Moreover, the spin nature of low-energy excitations is also manifested in their dispersions:
In the FM-LS phase where the spins develop ferromagnetic order, the Goldstone
mode has a quadratic dispersion, $\omega\propto k^{2}$,
as shown in Fig. \ref{fig6}(a);
In contrast, in the AFM-LS phase, the antiferromagnetic spin order gives rise to
two degenerate linearly dispersing Goldstone modes, $\omega\propto k$,
as shown in Fig. \ref{fig6}(b).
Meanwhile, the high-energy excitations are fully gapped (green lines in Fig. \ref{fig6}(a, b)) and well separated from the low-energy spin excitations.
These modes are predominantly associated with the layer pseudospin degrees of
freedom, and the presence of a finite gap is consistent with the breaking of
the discrete pseudospin $\mathbb{Z}_{2}$ symmetry in the layer-staggered phases.

The low-energy spin-wave modes should be directly accessible via inelastic
neutron scattering. By contrast, the gapped pseudospin excitations, which are
associated primarily with inter-layer charge fluctuations, are invisible to
neutron scattering. Instead, such layer-resolved charge excitations may be more
naturally probed by spectroscopic techniques sensitive to orbital and charge
dynamics, such as resonant inelastic X-ray scattering (RIXS) \cite{RIXS}.

For the AFM-LC phase, continuous symmetries in both the spin and pseudospin sectors are spontaneously broken.
Specifically, in the spin sector the antiferromagnetic order $\langle \mathbf{S}\rangle \neq 0$ breaks the $O(3)$ spin-rotational symmetry down to $O(2)$, leading to two branches of linearly dispersive Goldstone modes as gapless spin fluctuations.
Meanwhile, the in-plane $\langle\hat\tau^+\rangle\neq 0$ spontaneously breaks
the $U(1)$ layer-phase symmetry and gives rise to a third linear Goldstone mode as pseudospin fluctuations, as shown in Fig. \ref{fig6} (c). The two Goldstone modes have different velocities in general, as they are associated with two independent symmetries.

In addition to these well-separated branches, we identify a nearly flat band,
marked by the black lines in Fig. \ref{fig6}(a-c). This mode cannot be
classified as a purely spin or purely layer excitation.
Instead, it corresponds to the
composite spin and layer entangled excitation.
Remarkably, as system parameters are tuned, this nearly flat mode gradually
develops dispersion and softens.
When the gap of this hybrid excitation closes, the system enters the
SLE phase, indicating that this mode plays a central
role in driving the transition.

The resulting excitation spectrum in the SLE phase (Fig. \ref{fig6}(d))
highlights the distinctive character of the spin-layer-entangled ground
state. In the FM-LS, AFM-LS, and
the AFM-LC phases, the low-energy
excitations can be unambiguously classified according to their quantum
numbers. Notably, even in the AFM-LC phase, where both spin and pseudospin
sectors are gapless, the corresponding modes remain decoupled: each
Goldstone mode originates from the spontaneous breaking of either
spin-rotation symmetry or the layer-phase symmetry, and can therefore
be identified as a purely spin or purely pseudospin excitation.

In stark contrast, the low-energy excitations in the SLE phase are
intrinsically hybridized. Owing to the spin\textendash orbital locking
in the ground state, independent spin and pseudospin fluctuations
are suppressed, and the relevant low-energy degrees of freedom correspond
to collective rotations of the entangled composite order parameter.
As a result, the gapless Goldstone modes cannot be uniquely
classified as purely spin or purely pseudospin excitations. Instead,
they are associated with rigid rotations of $\mathcal{Q}^{\alpha\beta}$
within the emergent $S^3/\mathbb{Z}_2$ manifold, reflecting the locked nature
of spin and layer dynamics. The spontaneous symmetry breaking follows
the pattern $O(4)\rightarrow O(3)$, where the residual
$O(3)$ corresponds to the simultaneous rotation of
locked spin and pseudospin moments. This symmetry breaking naturally
accounts for the three gapless branches observed in the spectrum.

\section{Discussions and Conclusions}

In this work, we have theoretically investigated the low-energy physics
of a quarter-hole-filled bilayer Hubbard model, particularly focusing
on the role of orbital-selective interlayer hybridization. By explicitly
treating the strong vertical bonding of $d_{z^{2}}$ orbitals within
a molecular orbital basis and projecting out high-energy states,
we derived an effective anisotropic Kugel-Khomskii Hamiltonian that
describes the coupled dynamics of the
electron spin and an emergent layer
pseudospin associated with electrons in the
$d_{x^{2}-y^{2}}$ orbitals.
The resulting effective model is highly anisotropic in the pseudospin
space and exhibits a nontrivial interplay between spin exchange and
spin\textendash layer coupled interactions.

Using a combination of Weiss mean-field theory that fully retains
on-site quantum correlations between spin and layer degrees of freedom,
we established the ground state phase diagram characterized by four
distinct quantum phases. In addition to conventional magnetically
ordered phases accompanied by
layer-staggered electron occupation, we identified
a layer-coherent antiferromagnetic phase characterized by spontaneous
interlayer quantum coherence without static layer polarization. More
remarkably, we uncovered a spin\textendash layer\textendash entangled
(SLE) phase that does not exhibit any conventional dipolar order in
either the spin or layer sectors. Instead, this phase is characterized
by a hidden composite order parameter formed by the bilinear operators
$\hat{S}^{\beta}\hat{\tau}^{\alpha}$, signaling maximal local entanglement
between the two internal degrees of freedom.

We showed that the SLE phase originates from a strong and cooperative
coupling between the spin and layer degrees of freedom, leading to
the spontaneous breaking of an emergent O(4) symmetry.
This symmetry breaking gives rise to a distinctive excitation spectrum
featuring the entangled gapless Goldstone modes, which correspond
to collective rotations of a rigid spin\textendash layer-locked order
parameter rather than independent fluctuations of the individual sectors.

Recent experiments on the bilayer nickelate La$_{3}$Ni$_{2}$O$_{7}$
have
reported
signatures of possibly intertwined spin-density-wave
(SDW) and charge-density-wave (CDW) orders
\cite{li2025orbital,meng2024density,zhao2025pressure,meng2024normal}.
At ambient pressure, the low-temperature phase of La$_{3}$Ni$_{2}$O$_{7}$ adopts
the $Amam$ space group, and the resulting spin or charge pattern cannot be
stabilized within the unit cell of the minimal Kugel-Khomskii model we studied
in this work, which considers only nearest-neighbor hopping processes.
However, we expect the effective Kugel-Khomskii model is able to describe the
intertwined SDW and CDW order when additional effects, such as longer-range
hopping, residual itinerancy, and electron\textendash lattice coupling \cite{khasanov2025oxygen,yi2024nature,li2025direct,zhang2025spin,wang2025origin},
are taken into account.

Moreover, it is worth noting that the strong coupling between spin and
layer degrees of freedom revealed in our study may provide a natural setting
for the coexistence or mutual reinforcement of spin and charge ordering
tendencies.
In particular, ordering in the layer (pseudospin) sector necessarily involves
a modulation of electronic occupation between the two layers,
which can be viewed as a form of interlayer charge density modulation.
When combined with magnetic ordering in the spin sector, such layer-selective
charge redistribution may naturally accompany or enhance SDW instabilities,
leading to intertwined spin and charge ordering phenomena.
From this perspective, certain CDW signatures observed experimentally may not
originate from a conventional Peierls-type instability driven purely by Fermi
surface nesting, but could instead reflect charge modulations tied to
spin-layer correlations and orbital-selective interlayer hybridization.
While the minimal Kugel--Khomskii model considered here does not incorporate
the lattice anisotropy and spatially modulated interactions required to
stabilize SDW or CDW order at
the observed wave vectors, it suggests that spin and
charge degrees of freedom are intrinsically coupled in bilayer nickelates.
In more realistic settings that include structural distortions and longer-range
interactions, this intrinsic coupling may facilitate the emergence of
coexisting or intertwined SDW and CDW orders.

Beyond La$_{3}$Ni$_{2}$O$_{7}$, the mechanism identified in this
work is expected to be applicable to a broader class of correlated
systems. In particular, moir\'e bilayers and cold-atom optical lattices
offer highly tunable platforms in which layer, orbital, or valley
indices play roles analogous to the layer pseudospin considered here.
In these settings, interlayer tunneling, interaction anisotropy, and
filling can be controlled independently, providing promising opportunities
to engineer and probe spin\textendash layer entangled phases in a
controlled manner.
\section*{Acknowledgements}
We thank Y. Du for insightful discussions.  This work is supported by the National Key R\&D Program of China (Grant No.2023YFA1406500),
and the National Science Foundation of China (Grant Nos. 12334008, 12174441 and 12564021).
\appendix

\section*{Appendix}

\section{Schrieffer-Wolff transformation and Second perturbation}

To derive the effective model describing the exchange interactions
between these local moments, we employ the Schrieffer-Wolff transformation.
The effective Hamiltonian is generated via a unitary transformation
$H_{\text{eff}}=e^{s}He^{-s}$, which can be expanded as:
\begin{align}
  H_{\text{eff}} & =e^{s}He^{-s}=H_{0}+[S,\ H_{0}]+H^\prime+[S,\
  H^\prime]\label{eq:3}                                          \\
                 & +\frac{1}{2}[S,\ [S,\ H_{0}]]+...             %\nonumber
\end{align}
This formalism allows us to systematically eliminate the high-energy
degrees of freedom and construct an effective Hamiltonian $H_{\text{eff}}$
that acts solely within the low-energy subspace $\mathcal{S}$. The
unitary operator $S$ is determined by the condition that the first-order
terms vanish, i.e., $H^\prime+[H_{0},\ S]=0$. By solving for $S$ and substituting
it back into the Eq. \ref{eq:3}, the leading contribution to the
effective Hamiltonian is found to be of the second order,
$H_{\text{eff}}=\frac{1}{2}[S,\ H^\prime]$.

To evaluate the explicit form of the second-order terms, it is convenient
to introduce the projection operators $P$ and $Q=1-P$, which project
onto the low-energy manifold and the excited states, respectively.
Since the hopping amplitude $t_{\parallel}$ is much smaller than
the energy gap $\Delta E$ to the excited states, we can treat $H^\prime$
as a perturbation. The effective Hamiltonian up to the second order
is given:

%\[
\begin{equation}
  H_{\text{eff}}=PH^\prime P-PH^\prime Q\frac{1}{H_{0}-E_{0}}QH^\prime P
\end{equation}
%\]

Since the hopping term $H^\prime$ changes the local particle number, it
has no matrix element within the fixed-filling ground state manifold.
Consequently, the leading contribution arises from the second-order
term, which describes virtual hopping processes.

There are two major virtual hopping processes shown in Fig. \ref{fig3}.
In the Type-I process, the electrons at sites $i$ and $j$ reside
in the same layer ($\tau_{i,z}=\tau_{j,z}$) but possess antiparallel
spins. The hopping event generates a high-energy intermediate state
with on-site double occupancy. Subsequently, the system returns to
the low-energy subspace via a second in-plane hopping. Throughout
this process, the pseudospin configuration remains invariant ($\tau_{z}$
is conserved). However, the final spin arrangement has two possibilities:
the spins can either recover their original configuration or undergo
a spin exchange (spin-flip), as illustrated in Fig. \ref{fig3}(a).

In the Type-II process, the electrons at sites $i$ and $j$ initially
occupy different layers ($\tau_{i,z}\neq\tau_{j,z}$), while their
spin orientation can be arbitrary. The initial hopping generates an
intermediate state where the acceptor site accommodates electrons
in both the top and bottom layers. During the return hop, either of
the two electrons on the doubly occupied site can hop back to the
donor site. This leads to multiple possible outcomes: the system may
return to its original configuration, or undergo a spin flip, a pseudospin
flip, as illustrated in Fig. 3(b). For this process, we must address
the origin of the high excitation energy of the intermediate state
compared to the initial configuration. The primary reason lies in
the formation of a spin singlet within the doubly occupied bonding
$d_{z^{2}}$ orbital. When the $d_{x^{2}-y^{2}}$ orbitals are one
in each layer, the presence of this rigid $d_{z^{2}}$ singlet suppresses
the effective inter-orbital Hund's coupling. Unlike in a high-spin
atomic limit where electrons align to lower the energy, the $z_{2}$
singlet prevents ferromagnetic alignment with the $d_{x^{2}-y^{2}}$
electrons. Consequently, the system is unable to gain the Hund's correlation
energy, resulting in a significantly higher total energy.

\section{SU(4) flavor wave theory}

Given that the local Hilbert space has dimension $D_{l}=4$, we introduce
four Schwinger bosons (SBs) with annihilation (creation) operators
$b_{m,i}\ (b_{m,i}^{(\dagger)})$, with the flavor index $m\in\{0,1,2,3\}$.

The eigenstates of $S_{i}^{z}$ are expressed in terms of these bosons
as $b_{m,i}^{\dagger}|\emptyset\rangle=|m\rangle_{i}$. The local
constraint $\sum_{m=0}^{3}b_{m,i}^{\dagger}b_{m,i}=1$ projects the
bosonic operators onto the physical Hilbert space. To describe the
fluctuations above the ordered ground state obtained from our variational
calculation, we perform a site-dependent unitary transformation to
align the local quantization axis with the classical ground state
direction. We introduce a new set of bosonic operators $\tilde{b}_{n,i}$
($n\in\{0,1,2,3\}$), which are related to the original operators
$b_{m,i}$ via a local unitary matrix $\mathcal{U}_{i}$:
\begin{equation}
  b_{m,i}=\sum_{n=0}^{3}(\mathcal{U}_{i})_{mn}\tilde{b}_{n,i}.
\end{equation}

We then treat the system within the Holstein-Primakoff approximation.
Assuming the ground state is macroscopically occupied, we replace
the operator $\tilde{b}_{0,i}$ with a classical variable by condensing
the boson in the $n=0$ channel:
\begin{equation}
  \tilde{b}_{0,i}=\tilde{b}_{0,i}^{\dagger}
  \approx\sqrt{1-\sum_{n=1}^{3}\tilde{b}_{n,i}^{\dagger}\tilde{b}_{n,i}}.
\end{equation}

By substituting these expressions back into the original Hamiltonian
and retaining terms up to quadratic order, we obtain the spin-wave
Hamiltonian:

\begin{equation}
  H_{\text{SW}}=\frac{1}{2}\sum_{\mathbf{k}}
  \mathbf{\Psi}_{\mathbf{k}}^{\dagger}\mathcal{H}
  (\mathbf{k})\mathbf{\Psi}_{\mathbf{k}}+\text{const},
\end{equation}

where $\mathbf{\Psi}_{\mathbf{k}}$ is the Nambu spinor containing
the Fourier-transformed boson operators.

This quadratic Hamiltonian is diagonalized by a Bogoliubov transformation,
yielding the single-particle dispersions $E_{k\alpha}$:

\begin{equation}
  H_{\text{SW}}=\sum_{k,\alpha}E_{k\alpha}
  \left(a_{k\alpha}^{\dagger}a_{k\alpha}+\frac{1}{2}\right)+E_{0}.
\end{equation}

\bibliographystyle{apsrev4-2}
\bibliography{refs}

@article{Sun_Nature_2023,
  title     = {Signatures of superconductivity near 80 K in a nickelate under high pressure},
  author    = {Sun, Hualei and Huo, Mengwu and Hu, Xunwu and Li, Jingyuan and Liu, Zengjia and Han, Yifeng and Tang, Lingyun and Mao, Zhongquan and Yang, Pengtao and Wang, Bosen and others},
  journal   = {Nature},
  volume    = {621},
  number    = {7979},
  pages     = {493--498},
  year      = {2023},
  publisher = {Nature Publishing Group UK London}
}

@article{arXiv_2407_05681Cheng,
  title     = {Bulk high-temperature superconductivity in the high-pressure 
               tetragonal phase of bilayer La2PrNi2O7},
  author    = {Wang, Ningning and Wang, Gang and Shen, Xiaoling and Hou, Jun 
               and Luo, Jun and Ma, Xiaoping and Yang, Huaixin and Shi, Lifen and Dou, Jie 
               and Feng, Jie and Yang, Jie and Shi, Yunqing and Ren, Zhian and Ma, Hanming 
               and Yang, Pengtao and Liu, Ziyi and Liu, Yue and Zhang, Hua and Dong, 
               Xiaoli and Wang, Yuxin and Jiang, Kun and Hu, Jiangping and Nagasaki, Shoko 
               and Kitagawa, Kentaro and Calder, Stuart and Yan, Jiaqiang and Sun, 
               Jianping and Wang, Bosen and Zhou, Rui and Uwatoko, Yoshiya and Cheng, 
               Jinguang	},
  journal   = {Nature},
  volume    = {634},
  number    = {8034},
  pages     = {579},
  url       = {http://dx.doi.org/10.1038/s41586-024-07996-8},
  doi       = {10.1038/s41586-024-07996-8},
  year      = {2024},
  publisher = {Nature Publishing Group UK London}
}

@article{Yao_PRL_2023,
  title     = {Bilayer two-orbital model of La$_3$Ni$_2$O$_7$ under pressure},
  author    = {Luo, Zhihui and Hu, Xunwu and Wang, Meng and W{\'u}, W{\'e}i and Yao, Dao-Xin},
  journal   = {Phys. Rev. Lett},
  volume    = {131},
  number    = {12},
  pages     = {126001},
  year      = {2023},
  publisher = {APS}
}

@article{Liao_PRB_2023,
  title     = {Electron correlations and superconductivity in La$_3$Ni$_2$O$_7$ under pressure tuning},
  author    = {Liao, Zhiguang and Chen, Lei and Duan, Guijing and Wang, Yiming and Liu, Changle and Yu, Rong and Si, Qimiao},
  journal   = {Phys. Rev. B},
  volume    = {108},
  number    = {21},
  pages     = {214522},
  year      = {2023},
  publisher = {APS}
}

@article{ShilenkoLeonov_PRB_2023,
  title     = {Correlated electronic structure, orbital-selective behavior, and magnetic correlations in double-layer La 3 Ni 2 O 7 under pressure},
  author    = {Shilenko, DA and Leonov, IV},
  journal   = {Phys. Rev. B},
  volume    = {108},
  number    = {12},
  pages     = {125105},
  year      = {2023},
  publisher = {APS}
}

@article{LechermannEremin_PRB_2023,
  author    = {Lechermann, Frank and Gondolf, Jannik and B{\"o}tzel, Steffen and Eremin, Ilya M},
  journal   = {Phys. Rev. B},
  volume    = {108},
  number    = {20},
  pages     = {L201121},
  year      = {2023},
  publisher = {APS}
}

@article{CaoYang_PRB_2024,
  title     = {Flat bands promoted by Hund's rule coupling in the candidate double-layer high-temperature superconductor La 3 Ni 2 O 7 under high pressure},
  author    = {Cao, Yingying and Yang, Yi-feng},
  journal   = {Phys. Rev. B},
  volume    = {109},
  number    = {8},
  pages     = {L081105},
  year      = {2024},
  publisher = {APS}
}

@article{OuyangLu_PRB_2024,
  title     = {Hund electronic correlation in La 3 Ni 2 O 7 under high pressure},
  author    = {Ouyang, Zhenfeng and Wang, Jia-Ming and Wang, Jing-Xuan and He, Rong-Qiang and Huang, Li and Lu, Zhong-Yi},
  journal   = {Phys. Rev. B},
  volume    = {109},
  number    = {11},
  pages     = {115114},
  year      = {2024},
  publisher = {APS}
}

@article{RyeeWehling_PRL_2024,
  author    = {Ryee, Siheon and Witt, Niklas and Wehling, Tim O},
  journal   = {Phys. Rev. Lett},
  volume    = {133},
  number    = {9},
  pages     = {096002},
  year      = {2024},
  publisher = {APS}
}

@article{WangHu_arXiv_2024,
  title     = {Electronic and magnetic structures of bilayer La 3 Ni 2 O 7 at ambient pressure},
  author    = {Wang, Yuxin and Jiang, Kun and Wang, Ziqiang and Zhang, Fu-Chun and Hu, Jiangping},
  journal   = {Phys. Rev. B},
  volume    = {110},
  number    = {20},
  pages     = {205122},
  year      = {2024},
  publisher = {APS}
}

@article{QuSu_PRL_2024,
  title     = {Bilayer t-J-J$_\perp$ model and magnetically mediated pairing in the pressurized nickelate La 3 Ni 2 O 7},
  author    = {Qu, Xing-Zhou and Qu, Dai-Wei and Chen, Jialin and Wu, Congjun and Yang, Fan and Li, Wei and Su, Gang},
  journal   = {Phys. Rev. Lett},
  volume    = {132},
  number    = {3},
  pages     = {036502},
  year      = {2024},
  publisher = {APS}
}

@article{HeierSavrasov_PRB_2024,
  title   = {Competing $d_{xy}$ and $s_{\pm}$ pairing symmetries in superconducting {La$_3$Ni$_2$O$_7$} emerge from {LDA+FLEX} calculations},
  author  = {Heier, Griffin and Park, Kyungwha and Savrasov, Sergey Y.},
  journal = {Phys. Rev. B},
  volume  = {109},
  pages   = {104508},
  year    = {2024}
}

@article{ZhanHu_arXiv_2024,
  title     = {Cooperation between Electron-Phonon Coupling and Electronic 
               Interaction in Bilayer Nickelates 
               ${\mathrm{La}}_{3}{\mathrm{Ni}}_{2}{\mathrm{O}}_{7}$},
  author    = {Zhan, Jun and Gu, Yuhao and Wu, Xianxin and Hu, Jiangping},
  journal   = {Phys. Rev. Lett.},
  volume    = {134},
  issue     = {13},
  pages     = {136002},
  numpages  = {7},
  year      = {2025},
  month     = {Mar},
  publisher = {American Physical Society},
  doi       = {10.1103/PhysRevLett.134.136002},
  url       = {https://link.aps.org/doi/10.1103/PhysRevLett.134.136002}
}

@article{TianLu_PRB_2024,
  title     = {Correlation effects and concomitant two-orbital s$\pm$-wave superconductivity in La 3 Ni 2 O 7 under high pressure},
  author    = {Tian, Yi-Heng and Chen, Yin and Wang, Jia-Ming and He, Rong-Qiang and Lu, Zhong-Yi},
  journal   = {Phys. Rev. B},
  volume    = {109},
  number    = {16},
  pages     = {165154},
  year      = {2024},
  publisher = {APS}
}

@article{PanWu_arXiv_2023,
  title     = {Effect of rare-earth element substitution in superconducting R3Ni2O7 under pressure},
  author    = {Pan, Zhiming and Lu, Chen and Yang, Fan and Wu, Congjun},
  journal   = {Chin. Phys. Lett.},
  volume    = {41},
  number    = {8},
  pages     = {087401},
  year      = {2024},
  publisher = {IOP Publishing}
}

@article{ChangLi_arXiv_2023,
  title   = {Fermi surface symmetric mass generation: a quantum Monte-Carlo study},
  author  = {Chang, Wei-Xuan and Guo, Sibo and You, Yi-Zhuang and Li, Zi-Xiang},
  journal = {arXiv preprint arXiv:2311.09970},
  year    = {2023}
}

@article{WangYang_arXiv_2024,
  title   = {Highly asymmetric superconducting dome and strange metallicity in La$_3$Ni$_2$O$_7$},
  author  = {Wang, Jiangfan and Yang, Yi-feng },
  journal = {arXiv:2408.09774},
  year    = {2024}
}

@article{QinYang_PRB_2023,
  title     = {High-T c superconductivity by mobilizing local spin singlets and possible route to higher T c in pressurized La 3 Ni 2 O 7},
  author    = {Qin, Qiong and Yang, Yi-feng},
  journal   = {Phys. Rev. B},
  volume    = {108},
  number    = {14},
  pages     = {L140504},
  year      = {2023},
  publisher = {APS}
}

@article{LuoYao_npjQM_2024,
  title     = {High-T C superconductivity in La3Ni2O7 based on the bilayer two-orbital tJ model},
  author    = {Luo, Zhihui and Lv, Biao and Wang, Meng and W{\'u}, W{\'e}i and Yao, Dao-Xin},
  journal   = {npj Quantum Mater.},
  volume    = {9},
  number    = {1},
  pages     = {61},
  year      = {2024},
  publisher = {Nature Publishing Group UK London}
}

@article{JiangZhang_CPL_2024,
  title     = {High-temperature superconductivity in La$_3$Ni$_2$O$_7$},
  author    = {Jiang, Kun and Wang, Ziqiang and Zhang, Fu-Chun},
  journal   = {Chin. Phys. Lett.},
  volume    = {41},
  number    = {1},
  pages     = {017402},
  year      = {2024},
  publisher = {IOP Publishing}
}

@article{HuangZhou_PRB_2023,
  title     = {Impurity and vortex states in the bilayer high-temperature superconductor La 3 Ni 2 O 7},
  author    = {Huang, Junkang and Wang, ZD and Zhou, Tao},
  journal   = {Phys. Rev. B},
  volume    = {108},
  number    = {17},
  pages     = {174501},
  year      = {2023},
  publisher = {APS}
}

@article{YangZhang_PRB_2023,
  title     = {Interlayer valence bonds and two-component theory for high-T c superconductivity of La 3 Ni 2 O 7 under pressure},
  author    = {Yang, Yi-feng and Zhang, Guang-Ming and Zhang, Fu-Chun},
  journal   = {Phys. Rev. B},
  volume    = {108},
  number    = {20},
  pages     = {L201108},
  year      = {2023},
  publisher = {APS}
}

@article{LuWu_PRL_2024,
  title     = {Interlayer-coupling-driven high-temperature superconductivity in La 3 Ni 2 O 7 under pressure},
  author    = {Lu, Chen and Pan, Zhiming and Yang, Fan and Wu, Congjun},
  journal   = {Phys. Rev. Lett},
  volume    = {132},
  number    = {14},
  pages     = {146002},
  year      = {2024},
  publisher = {APS}
}

@article{LuWu_PRB_2024,
  title     = {Interplay of two E$_$g orbitals in superconducting La 3 Ni 2 O 7 under pressure},
  author    = {Lu, Chen and Pan, Zhiming and Yang, Fan and Wu, Congjun},
  journal   = {Phys. Rev. B},
  volume    = {110},
  number    = {9},
  pages     = {094509},
  year      = {2024},
  publisher = {APS}
}

@article{XueWang_CPL_2024,
  title     = {Magnetism and Superconductivity in the t--J Model of La3Ni2O7 Under Multiband Gutzwiller Approximation},
  author    = {Xue, Jie-Ran and Wang, Fa},
  journal   = {Chin. Phys. Lett.},
  volume    = {41},
  number    = {5},
  pages     = {057403},
  year      = {2024},
  publisher = {IOP Publishing}
}

@article{ChenLi_PRB_2024,
  title     = {Orbital-selective superconductivity in the pressurized bilayer nickelate La 3 Ni 2 O 7: An infinite projected entangled-pair state study},
  author    = {Chen, Jialin and Yang, Fan and Li, Wei},
  journal   = {Phys. Rev. B},
  volume    = {110},
  number    = {4},
  pages     = {L041111},
  year      = {2024},
  publisher = {APS}
}

@article{KanekoKuroki_PRB_2024,
  title     = {Pair correlations in the two-orbital Hubbard ladder: Implications for superconductivity in the bilayer nickelate La 3 Ni 2 O 7},
  author    = {Kaneko, Tatsuya and Sakakibara, Hirofumi and Ochi, Masayuki and Kuroki, Kazuhiko},
  journal   = {Phys. Rev. B},
  volume    = {109},
  number    = {4},
  pages     = {045154},
  year      = {2024},
  publisher = {APS}
}

@article{KakoiKuroki_PRB_2024,
  title     = {Pair correlations of the hybridized orbitals in a ladder model for the bilayer nickelate La 3 Ni 2 O 7},
  author    = {Kakoi, Masataka and Kaneko, Tatsuya and Sakakibara, Hirofumi and Ochi, Masayuki and Kuroki, Kazuhiko},
  journal   = {Phys. Rev. B},
  volume    = {109},
  number    = {20},
  pages     = {L201124},
  year      = {2024},
  publisher = {APS}
}

@article{MaWu_arXiv_2024,
  title   = {Parameters dependent superconducting transition temperature in high temperature superconductors},
  author  = {Ma, Runyu and Ma, Tianxing and Wu, Congjun},
  journal = {arXiv preprint arXiv:2408.02031},
  year    = {2024}
}

@article{SakakibaraKuroki_PRL_2024,
  title     = {Possible high T c superconductivity in La 3 Ni 2 O 7 under high pressure through manifestation of a nearly half-filled bilayer Hubbard model},
  author    = {Sakakibara, Hirofumi and Kitamine, Naoya and Ochi, Masayuki and Kuroki, Kazuhiko},
  journal   = {Phys. Rev. Lett},
  volume    = {132},
  number    = {10},
  pages     = {106002},
  year      = {2024},
  publisher = {APS}
}

@article{YangWang_PRB_2023,
  author  = {Yang, Qing-Geng and Wang, Da and Wang, Qiang-Hua },
  title   = {Possible s$_{\pm}$-wave superconductivity in La$_3$Ni$_2$O$_7$},
  journal = {Phys. Rev. B},
  volume  = {108},
  pages   = {L140505},
  year    = {2023}
}

@article{JiangKu_PRL_2024,
  title     = {Pressure driven fractionalization of ionic spins results in cupratelike high-T c superconductivity in La 3 Ni 2 O 7},
  author    = {Jiang, Ruoshi and Hou, Jinning and Fan, Zhiyu and Lang, Zi-Jian and Ku, Wei},
  journal   = {Phys. Rev. Lett},
  volume    = {132},
  number    = {12},
  pages     = {126503},
  year      = {2024},
  publisher = {APS}
}

@article{LiuChen_arXiv_2023,
  title     = {Role of crystal-field-splitting and longe-range-hoppings on superconducting pairing symmetry of La $ \_3 $ Ni $ \_2 $ O $ \_7$},
  author    = {Liu, Hongquan and Xia, Chengliang and Zhou, Shengjie and Chen, Hanghui},
  journal   = {Nature Communications},
  volume    = {16},
  number    = {1},
  pages     = {1054},
  url       = {http://dx.doi.org/10.1038/s41467-025-56206-0},
  doi       = {10.1038/s41467-025-56206-0},
  year      = {2025},
  publisher = {Nature Publishing Group UK London}
}

@article{QuSu_arXiv_2023,
  title     = {Hund's rule, interorbital hybridization, and high-${T}_{c}$ 
               superconductivity in the bilayer nickelate 
               ${\mathrm{La}}_{3}{\mathrm{Ni}}_{2}{\mathrm{O}}_{7}$},
  author    = {Qu, Xing-Zhou and Qu, Dai-Wei and Yi, Xin-Wei and Li, Wei and Su, 
               Gang},
  journal   = {Phys. Rev. B},
  volume    = {112},
  issue     = {16},
  pages     = {L161101},
  numpages  = {8},
  year      = {2025},
  month     = {Oct},
  publisher = {American Physical Society},
  doi       = {10.1103/171w-6kjw},
  url       = {https://link.aps.org/doi/10.1103/171w-6kjw}
}

@article{YangZhang_arXiv_2024,
  title   = {Strong pairing and symmetric pseudogap metal in double Kondo lattice model: from nickelate superconductor to tetralayer optical lattice},
  author  = {Yang, Hui and Oh, Hanbit and Zhang, Ya-Hui},
  journal = {arXiv preprint arXiv:2408.01493},
  year    = {2024}
}

@article{YangZhang_arXiv_2023,
  title     = {Strong pairing from a small Fermi surface beyond weak coupling: Application to La 3 Ni 2 O 7},
  author    = {Yang, Hui and Oh, Hanbit and Zhang, Ya-Hui},
  journal   = {Phys. Rev. B},
  volume    = {110},
  number    = {10},
  pages     = {104517},
  year      = {2024},
  publisher = {APS}
}

@article{ZhangWeng_PRL_2024,
  title     = {Strong pairing originated from an emergent Z 2 Berry phase in La 3 Ni 2 O 7},
  author    = {Zhang, Jia-Xin and Zhang, Hao-Kai and You, Yi-Zhuang and Weng, Zheng-Yu},
  journal   = {Phys. Rev. Lett},
  volume    = {133},
  number    = {12},
  pages     = {126501},
  year      = {2024},
  publisher = {APS}
}

@article{LuYou_arXiv_2023,
  title   = {Superconductivity from Doping Symmetric Mass Generation Insulators: Application to La $ \_3 $ Ni $ \_2 $ O $ \_7 $ under Pressure},
  author  = {Lu, Da-Chuan and Li, Miao and Zeng, Zhao-Yi and Hou, Wanda and Wang, Juven and Yang, Fan and You, Yi-Zhuang},
  journal = {arXiv preprint arXiv:2308.11195},
  year    = {2023}
}

@article{FanXiang_PRB_2024,
  title     = {Superconductivity in nickelate and cuprate superconductors with strong bilayer coupling},
  author    = {Fan, Zhen and Zhang, Jian-Feng and Zhan, Bo and Lv, Dingshun and Jiang, Xing-Yu and Normand, Bruce and Xiang, Tao},
  journal   = {Phys. Rev. B},
  volume    = {110},
  number    = {2},
  pages     = {024514},
  year      = {2024},
  publisher = {APS}
}

@article{ZhengWu_arXiv_2023,
  title     = {s$\pm$-wave superconductivity in the bilayer two-orbital Hubbard model},
  author    = {Zheng, Yao-Yuan and W{\'u}, W{\'e}i},
  journal   = {Phys. Rev. B},
  volume    = {111},
  number    = {3},
  pages     = {035108},
  year      = {2025},
  publisher = {APS}
}

@article{SchlomerBohrdt_arXiv_2023,
  title     = {Superconductivity in the pressurized nickelate La3Ni2O7 in the vicinity of a BEC--BCS crossover},
  author    = {Schl{\"o}mer, Henning and Schollw{\"o}ck, Ulrich and Grusdt, Fabian and Bohrdt, Annabelle},
  journal   = {Commun. Phys.},
  volume    = {7},
  number    = {1},
  pages     = {366},
  year      = {2024},
  publisher = {Nature Publishing Group UK London}
}

@article{BotzelEremin_arXiv_2024_1,
  title     = {Theory of magnetic excitations in the multilayer nickelate superconductor La 3 Ni 2 O 7},
  author    = {B{\"o}tzel, Steffen and Lechermann, Frank and Gondolf, Jannik and Eremin, Ilya M},
  journal   = {Phys. Rev. B},
  volume    = {109},
  number    = {18},
  pages     = {L180502},
  year      = {2024},
  publisher = {APS}
}

@article{OhZhang_arXiv_2024,
  title   = {Type II tJ model in charge transfer regime in bilayer La $ \_3 $ Ni $ \_2 $ O $ \_7 $ and trilayer La $ \_4 $ Ni $ \_3 $ O $ \_ $\{$10$\}$ $},
  author  = {Oh, Hanbit and Zhou, Boran and Zhang, Ya-Hui},
  journal = {arXiv preprint arXiv:2405.00092},
  year    = {2024}
}

@article{liao2412,
  title   = {Orbital-selective electron correlations in high-$ T\_ $\$$\backslash$rm c$\$ $ bilayer nickelates: from a global phase diagram to implications for spectroscopy},
  author  = {Liao, Zhiguang and Wang, Yiming and Chen, Lei and Duan, Guijing and Yu, Rong and Si, Qimiao},
  journal = {arXiv preprint arXiv:2412.21019},
  year    = {2024}
}

@article{yang2025film1,
  title     = {Origin of the diagonal double-stripe spin density wave and 
               potential superconductivity in bulk 
               ${\mathrm{La}}_{3}{\mathrm{Ni}}_{2}{\mathrm{O}}_{7}$ at ambient pressure},
  author    = {Liu, Yu-Bo and Sun, Hongyi and Zhang, Ming and Liu, Qihang and 
               Chen, Wei-Qiang and Yang, Fan},
  journal   = {Phys. Rev. B},
  volume    = {112},
  issue     = {1},
  pages     = {014510},
  numpages  = {13},
  year      = {2025},
  month     = {Jul},
  publisher = {American Physical Society},
  doi       = {10.1103/24f4-349n},
  url       = {https://link.aps.org/doi/10.1103/24f4-349n}
}

@article{hu2025film,
  title   = {Landscape of Correlated Orders in Strained Bilayer Nickelate Thin Films},
  author  = {Le, Congcong and Zhan, Jun and Wu, Xianxin and Hu, Jiangping},
  journal = {arXiv preprint arXiv:2501.14665},
  year    = {2025}
}

@article{ZhangDagotto_PRB_2023,
  title     = {Electronic structure, dimer physics, orbital-selective behavior, and magnetic tendencies in the bilayer nickelate superconductor ${\mathrm{La}}_{3}{\mathrm{Ni}}_{2}{\mathrm{O}}_{7}$ under pressure},
  author    = {Zhang, Yang and Lin, Ling-Fang and Moreo, Adriana and Dagotto, Elbio},
  journal   = {Phys. Rev. B},
  volume    = {108},
  issue     = {18},
  pages     = {L180510},
  numpages  = {5},
  year      = {2023},
  month     = {Nov},
  publisher = {American Physical Society},
  doi       = {10.1103/PhysRevB.108.L180510}
}

@article{ZhangDagotto_NC_2024,
  title     = { Structural phase transition, s$\pm$-wave pairing, and magnetic stripe order in bilayered superconductor La$_3$Ni$_2$O$_7$ under pressure },
  author    = {Zhang, Yang and Lin, Ling-Fang and Moreo, Adriana and Maier, Thomas A. and Dagotto, Elbio},
  journal   = { Nat. Commun. },
  volume    = {15},
  number    = {1},
  pages     = {2470},
  year      = {2024},
  month     = {Mar},
  publisher = { Nature Publishing Group UK London },
  doi       = {10.1038/s41467-024-46622-z }
}

@article{paschen2021quantum,
  title     = {Quantum phases driven by strong correlations},
  author    = {Paschen, Silke and Si, Qimiao},
  journal   = {Nature Reviews Physics},
  volume    = {3},
  number    = {1},
  pages     = {9--26},
  year      = {2021},
  publisher = {Nature Publishing Group UK London}
}

@article{schaffer2016recent,
  title     = {Recent progress on correlated electron systems with strong spin--orbit coupling},
  author    = {Schaffer, Robert and Lee, Eric Kin-Ho and Yang, Bohm-Jung and Kim, Yong Baek},
  journal   = {Reports on Progress in Physics},
  volume    = {79},
  number    = {9},
  pages     = {094504},
  year      = {2016},
  publisher = {IOP Publishing}
}

@article{liu2021orbital,
  title     = {Orbital magnetic states in moir{\'e} graphene systems},
  author    = {Liu, Jianpeng and Dai, Xi},
  journal   = {Nature Reviews Physics},
  volume    = {3},
  number    = {5},
  pages     = {367--382},
  year      = {2021},
  publisher = {Nature Publishing Group UK London}
}

@article{TMD2011,
  title     = {Tunable band gaps in bilayer transition-metal dichalcogenides},
  author    = {Ramasubramaniam, Ashwin and Naveh, Doron and Towe, Elias},
  journal   = {Physical Review B},
  volume    = {84},
  number    = {20},
  pages     = {205325},
  year      = {2011},
  publisher = {APS}
}

@article{zhang2020flat,
  title     = {Flat bands in twisted bilayer transition metal dichalcogenides},
  author    = {Zhang, Zhiming and Wang, Yimeng and Watanabe, Kenji and Taniguchi, Takashi and Ueno, Keiji and Tutuc, Emanuel and LeRoy, Brian J},
  journal   = {Nature Physics},
  volume    = {16},
  number    = {11},
  pages     = {1093--1096},
  year      = {2020},
  publisher = {Nature Publishing Group UK London}
}

@article{dai2016twisted,
  title     = {Twisted bilayer graphene: Moir{\'e} with a twist},
  author    = {Dai, Shuyang and Xiang, Yang and Srolovitz, David J},
  journal   = {Nano letters},
  volume    = {16},
  number    = {9},
  pages     = {5923--5927},
  year      = {2016},
  publisher = {ACS Publications}
}

@article{kariyado2023twisted,
  title     = {Twisted bilayer BC 3: Valley interlocked anisotropic flat bands},
  author    = {Kariyado, Toshikaze},
  journal   = {Physical Review B},
  volume    = {107},
  number    = {8},
  pages     = {085127},
  year      = {2023},
  publisher = {APS}
}

@article{chenKK,
  title     = {Emergent Berezinskii-Kosterlitz-Thouless and Kugel-Khomskii physics in the triangular lattice bilayer colbaltate},
  author    = {Chen, Gang V},
  journal   = {Physical Review Letters},
  volume    = {133},
  number    = {13},
  pages     = {136703},
  year      = {2024},
  publisher = {APS}
}

@article{chenwu2024,
  title     = {Multiflavor Mott insulators in quantum materials and ultracold atoms},
  author    = {Chen, Gang V and Wu, Congjun},
  journal   = {npj Quantum Materials},
  volume    = {9},
  number    = {1},
  pages     = {1},
  year      = {2024},
  publisher = {Nature Publishing Group UK London}
}

@article{dagotto2005,
  title     = {Complexity in strongly correlated electronic systems},
  author    = {Dagotto, Elbio},
  journal   = {Science},
  volume    = {309},
  number    = {5732},
  pages     = {257--262},
  year      = {2005},
  publisher = {American Association for the Advancement of Science}
}

@article{KKM,
  title     = {The Jahn-Teller effect and magnetism: transition metal compounds},
  author    = {Kugel, Kliment I and Khomski{\u\i}, Dmitri I},
  journal   = {Soviet Physics Uspekhi},
  volume    = {25},
  number    = {4},
  pages     = {231},
  year      = {1982},
  publisher = {IOP Publishing}
}

@article{streltsov2017orbital,
  title     = {Orbital physics in transition metal compounds: new trends},
  author    = {Streltsov, Sergey V and Khomskii, Daniel I},
  journal   = {Physics-Uspekhi},
  volume    = {60},
  number    = {11},
  pages     = {1121},
  year      = {2017},
  publisher = {IOP Publishing}
}

@article{brzezicki2011entangled,
  title     = {Entangled spin-orbital phases in the bilayer Kugel-Khomskii model},
  author    = {Brzezicki, Wojciech and Ole{\'s}, Andrzej M},
  journal   = {Physical Review B—Condensed Matter and Materials Physics},
  volume    = {83},
  number    = {21},
  pages     = {214408},
  year      = {2011},
  publisher = {APS}
}

@article{zhang2018lowest,
  title   = {Lowest-energy Moir$\backslash$'e Band Formed by Dirac Zero Modes in Twisted Bilayer Graphene},
  author  = {Zhang, Long},
  journal = {arXiv preprint arXiv:1804.09047},
  year    = {2018}
}

@article{khaliullin1997spin,
  title     = {Spin and orbital excitation spectrum in the Kugel-Khomskii model},
  author    = {Khaliullin, G and Oudovenko, V},
  journal   = {Physical Review B},
  volume    = {56},
  number    = {22},
  pages     = {R14243},
  year      = {1997},
  publisher = {APS}
}

@article{brzezicki2013exotic,
  title     = {Exotic spin orders driven by orbital fluctuations in the Kugel-Khomskii model},
  author    = {Brzezicki, Wojciech and Dziarmaga, Jacek and Ole{\'s}, Andrzej M},
  journal   = {Physical Review B—Condensed Matter and Materials Physics},
  volume    = {87},
  number    = {6},
  pages     = {064407},
  year      = {2013},
  publisher = {APS}
}

@article{kugel2015spin,
  title     = {Spin-orbital interaction for face-sharing octahedra: Realization of a highly symmetric SU (4) model},
  author    = {Kugel, KI and Khomskii, DI and Sboychakov, AO and Streltsov, SV},
  journal   = {Physical Review B},
  volume    = {91},
  number    = {15},
  pages     = {155125},
  year      = {2015},
  publisher = {APS}
}

@article{wang2009z,
  title     = {Z 2 spin-orbital liquid state in the square lattice Kugel-Khomskii model},
  author    = {Wang, Fa and Vishwanath, Ashvin},
  journal   = {Physical Review B—Condensed Matter and Materials Physics},
  volume    = {80},
  number    = {6},
  pages     = {064413},
  year      = {2009},
  publisher = {APS}
}

@article{calvera2021theory,
  title   = {Theory of Dirac Spin-Orbital Liquids: monopoles, anomalies, and applications to $ SU (4) $ honeycomb models},
  author  = {Calvera, Vladimir and Wang, Chong},
  journal = {arXiv preprint arXiv:2103.13405},
  year    = {2021}
}

@article{zhang2024variational,
  title     = {Variational Monte Carlo approach to the SU (4) spin-orbital model on the triangular lattice},
  author    = {Zhang, Chun and Jin, Hui-Ke and Zhou, Yi},
  journal   = {Physical Review B},
  volume    = {109},
  number    = {12},
  pages     = {125103},
  year      = {2024},
  publisher = {APS}
}

@article{gorshkov2010two,
  title     = {Two-orbital SU (N) magnetism with ultracold alkaline-earth atoms},
  author    = {Gorshkov, Alexey Vyacheslavovich and Hermele, M and Gurarie, V and Xu, C and Julienne, Paul S and Ye, J and Zoller, Peter and Demler, Eugene and Lukin, Mikhail D and Rey, AM},
  journal   = {Nature physics},
  volume    = {6},
  number    = {4},
  pages     = {289--295},
  year      = {2010},
  publisher = {Nature Publishing Group UK London}
}

@article{jin2023twisting,
  title     = {Twisting the Dirac cones of the SU (4) spin-orbital liquid on the honeycomb lattice},
  author    = {Jin, Hui-Ke and Natori, WMH and Knolle, Johannes},
  journal   = {Physical Review B},
  volume    = {107},
  number    = {18},
  pages     = {L180401},
  year      = {2023},
  publisher = {APS}
}

@article{gotfryd2020spin,
  title     = {How spin-orbital entanglement depends on the spin-orbit coupling in a Mott insulator},
  author    = {Gotfryd, Dorota and P{\"a}rschke, Ekaterina M and Chaloupka, Ji{\v{r}}{\'\i} and Ole{\'s}, Andrzej M and Wohlfeld, Krzysztof},
  journal   = {Physical Review Research},
  volume    = {2},
  number    = {1},
  pages     = {013353},
  year      = {2020},
  publisher = {APS}
}

@article{carr2020electronic,
  title     = {Electronic-structure methods for twisted moir{\'e} layers},
  author    = {Carr, Stephen and Fang, Shiang and Kaxiras, Efthimios},
  journal   = {Nature Reviews Materials},
  volume    = {5},
  number    = {10},
  pages     = {748--763},
  year      = {2020},
  publisher = {Nature Publishing Group UK London}
}

@article{cao2020tunable,
  title     = {Tunable correlated states and spin-polarized phases in twisted bilayer--bilayer graphene},
  author    = {Cao, Yuan and Rodan-Legrain, Daniel and Rubies-Bigorda, Oriol and Park, Jeong Min and Watanabe, Kenji and Taniguchi, Takashi and Jarillo-Herrero, Pablo},
  journal   = {Nature},
  volume    = {583},
  number    = {7815},
  pages     = {215--220},
  year      = {2020},
  publisher = {Nature Publishing Group UK London}
}

@article{cao2016superlattice,
  title     = {Superlattice-induced insulating states and valley-protected orbits in twisted bilayer graphene},
  author    = {Cao, Yuan and Luo, JY and Fatemi, Valla and Fang, Shiang and Sanchez-Yamagishi, JD and Watanabe, Kenji and Taniguchi, Takashi and Kaxiras, Efthimios and Jarillo-Herrero, Pablo},
  journal   = {Physical review letters},
  volume    = {117},
  number    = {11},
  pages     = {116804},
  year      = {2016},
  publisher = {APS}
}

@article{chu2020review,
  title     = {A review of experimental advances in twisted graphene moir{\'e} superlattice},
  author    = {Chu, Yanbang and Liu, Le and Yuan, Yalong and Shen, Cheng and Yang, Rong and Shi, Dongxia and Yang, Wei and Zhang, Guangyu},
  journal   = {Chinese Physics B},
  volume    = {29},
  number    = {12},
  pages     = {128104},
  year      = {2020},
  publisher = {IOP Publishing}
}

@article{li2025identification,
  title     = {Identification of superconductivity in bilayer nickelate La3Ni2O7 under high pressure up to 100 GPa},
  author    = {Li, Jingyuan and Peng, Di and Ma, Peiyue and Zhang, Hengyuan and Xing, Zhenfang and Huang, Xing and Huang, Chaoxin and Huo, Mengwu and Hu, Deyuan and Dong, Zixian and others},
  journal   = {National Science Review},
  pages     = {nwaf220},
  year      = {2025},
  publisher = {Oxford University Press}
}

@article{ji2025strong,
  title   = {A Strong-Coupling-Limit Study on the Pairing Mechanism in the Pressurized La $ \_3 $ Ni $ \_2 $ O $ \_7$},
  author  = {Ji, Jia-Heng and Lu, Chen and Shao, Zhi-Yan and Pan, Zhiming and Yang, Fan and Wu, Congjun},
  journal = {arXiv preprint arXiv:2504.12127},
  year    = {2025}
}

@article{kaneko2025t,
  title     = {t-J model for strongly correlated two-orbital systems: Application to bilayer nickelate superconductors},
  author    = {Kaneko, Tatsuya and Kakoi, Masataka and Kuroki, Kazuhiko},
  journal   = {Physical Review B},
  volume    = {112},
  number    = {7},
  pages     = {075143},
  year      = {2025},
  publisher = {APS}
}

@article{schrieffer1966relation,
  title     = {Relation between the anderson and kondo hamiltonians},
  author    = {Schrieffer, John R and Wolff, Peter A},
  journal   = {Physical Review},
  volume    = {149},
  number    = {2},
  pages     = {491},
  year      = {1966},
  publisher = {APS}
}

@article{anderson1950antiferromagnetism,
  title     = {Antiferromagnetism. Theory of superexchange interaction},
  author    = {Anderson, Philip W},
  journal   = {Physical Review},
  volume    = {79},
  number    = {2},
  pages     = {350},
  year      = {1950},
  publisher = {APS}
}

@article{li2025orbital,
  title   = {Orbital Signatures of Density Wave Transition in La3Ni2O7-delta and La2PrNi2O7-delta RP-Nickelates Probed via in-situ X-ray Absorption Near-edge Spectroscopy},
  author  = {Li, Mingtao and Zhang, Mingxin and Wang, Yiming and Guan, Jiayi and Li, Nana and Pei, Cuiying and Adama, N and Kong, Qingyu and Qi, Yanpeng and Yang, Wenge and others},
  journal = {arXiv preprint arXiv:2502.10962},
  year    = {2025}
}

@article{khasanov2025oxygen,
  title   = {Oxygen-isotope effect on the density wave transitions in La $ \_3 $ Ni $ \_2 $ O $ \_ $\{$7$\}$ $ and La $ \_4 $ Ni $ \_3 $ O $ \_ $\{$10$\}$ $},
  author  = {Khasanov, Rustem and Sazgari, Vahid and Plokhikh, Igor and Medarde, Marisa and Pomjakushina, Ekaterina and Klimczuk, Tomasz and Kr{\u{A}}{\l}lak, Szymon and Winiarski, Micha{\'L} and Hicken, Thomas J and Luetkens, Hubertus and others},
  journal = {arXiv preprint arXiv:2504.08290},
  year    = {2025}
}

@article{yi2024nature,
  title     = {Nature of charge density waves and metal-insulator transition in pressurized La 3 Ni 2 O 7},
  author    = {Yi, Xin-Wei and Meng, Ying and Li, Jia-Wen and Liao, Zheng-Wei and Li, Wei and You, Jing-Yang and Gu, Bo and Su, Gang},
  journal   = {Physical Review B},
  volume    = {110},
  number    = {14},
  pages     = {L140508},
  year      = {2024},
  publisher = {APS}
}

@article{li2025direct,
  title     = {Direct visualization of an incommensurate unidirectional charge density wave in L a 4 N i 3 O 10},
  author    = {Li, Mingzhe and Gong, Jiashuo and Zhu, Yinghao and Chen, Ziyuan and Zhang, Jiakang and Zhang, Enkang and Li, Yuanji and Yin, Ruotong and Wang, Shiyuan and Zhao, Jun and others},
  journal   = {Physical Review B},
  volume    = {112},
  number    = {4},
  pages     = {045132},
  year      = {2025},
  publisher = {APS}
}

@article{meng2024density,
  title     = {Density-wave-like gap evolution in La3Ni2O7 under high pressure revealed by ultrafast optical spectroscopy},
  author    = {Meng, Yanghao and Yang, Yi and Sun, Hualei and Zhang, Sasa and Luo, Jianlin and Chen, Liucheng and Ma, Xiaoli and Wang, Meng and Hong, Fang and Wang, Xinbo and others},
  journal   = {Nature Communications},
  volume    = {15},
  number    = {1},
  pages     = {10408},
  year      = {2024},
  publisher = {Nature Publishing Group UK London}
}

@article{zhang2025spin,
  title     = {Spin-charge-orbital order in nickelate superconductors},
  author    = {Zhang, Binhua and Xu, Changsong and Xiang, Hongjun},
  journal   = {Physical Review B},
  volume    = {111},
  number    = {18},
  pages     = {184401},
  year      = {2025},
  publisher = {APS}
}

@article{zhao2025pressure,
  title     = {Pressure-enhanced spin-density-wave transition in double-layer nickelate La3Ni2O7- $\delta$},
  author    = {Zhao, Dan and Zhou, Yanbing and Huo, Mengwu and Wang, Yu and Nie, Linpeng and Yang, Ye and Ying, Jianjun and Wang, Meng and Wu, Tao and Chen, Xianhui},
  journal   = {Science Bulletin},
  year      = {2025},
  publisher = {Elsevier}
}

@article{meng2024normal,
  title   = {Normal and Superconducting Properties of La 3 Ni 2 O 7},
  author  = {Meng, Wang and Hai-Hu, Wen and Tao, Wu and Dao-Xin, Yao and Tao, Xiang},
  journal = {Chin. Phys. Lett.},
  volume  = {41},
  number  = {7},
  year    = {2024}
}

@article{cui2024strain,
  title     = {Strain-mediated phase crossover in Ruddlesden--Popper nickelates},
  author    = {Cui, Ting and Choi, Songhee and Lin, Ting and Liu, Chen and Wang, Gang and Wang, Ningning and Chen, Shengru and Hong, Haitao and Rong, Dongke and Wang, Qianying and others},
  journal   = {Communications Materials},
  volume    = {5},
  number    = {1},
  pages     = {32},
  year      = {2024},
  publisher = {Nature Publishing Group UK London}
}

@article{duan2025orbital,
  title   = {Orbital-selective correlation effects and superconducting pairing symmetry in a multiorbital $ t $-$ J $ model for bilayer nickelates},
  author  = {Duan, Guijing and Liao, Zhiguang and Chen, Lei and Wang, Yiming and Yu, Rong and Si, Qimiao},
  journal = {arXiv preprint arXiv:2502.09195},
  year    = {2025}
}

@article{dahlbom2025sunny,
  title   = {Sunny. jl: a Julia package for spin dynamics},
  author  = {Dahlbom, David and Zhang, Hao and Miles, Cole and Quinn, Sam and Niraula, Alin and Thipe, Bhushan and Wilson, Matthew and Matin, Sakib and Mankad, Het and Hahn, Steven and others},
  journal = {arXiv preprint arXiv:2501.13095},
  year    = {2025}
}

@article{wang2025origin,
  title   = {Origin of Spin Stripes in Bilayer Nickelate La $ \_3 $ Ni $ \_2 $ O $ \_7$},
  author  = {Wang, Hao-Xin and Oh, Hanbit and Helbig, Tobias and Wang, Bai Yang and Li, Jiarui and Yu, Yijun and Hwang, Harold Y and Jiang, Hong-Chen and Wu, Yi-Ming and Raghu, S},
  journal = {arXiv preprint arXiv:2509.25344},
  year    = {2025}
}

@article{georges2013strong,
  title     = {Strong correlations from Hund’s coupling},
  author    = {Georges, Antoine and Medici, Luca de' and Mravlje, Jernej},
  journal   = {Annu. Rev. Condens. Matter Phys.},
  volume    = {4},
  number    = {1},
  pages     = {137--178},
  year      = {2013},
  publisher = {Annual Reviews}
}

@article{RIXS,
  title   = {Witnessing Spin-Orbital Entanglement using Resonant Inelastic X-Ray Scattering},
  author  = {Shen, Zecheng and Ding, Shuhan and Zhao, Zijun and Evangelista, Francesco A and Wang, Yao},
  journal = {arXiv preprint arXiv:2512.06718},
  year    = {2025}
}
\end{document}